\documentclass[iop]{emulateapj}
\usepackage{psfig}
\usepackage{apjfonts}
\usepackage{ulem}
\usepackage[dvips]{color}
\usepackage{graphicx}
\usepackage[USenglish]{babel}
\usepackage{mathrsfs}

\begin{document}

\title{UV INSIGHTS INTO THE COMPLEX POPULATIONS OF M~87 GLOBULAR
  CLUSTERS$^{\ast}$ }

\author{A.\ Bellini\altaffilmark{1},
A.\ Renzini\altaffilmark{2},
J.\ Anderson\altaffilmark{1},
L.\ R.\ Bedin\altaffilmark{2},
G.\ Piotto\altaffilmark{2,3},
M.\ Soto\altaffilmark{1},
T.\ M.\ Brown\altaffilmark{1},
A.\ P.\ Milone\altaffilmark{4},
S.\ T.\ Sohn\altaffilmark{5},
A.\ V.\ Sweigart\altaffilmark{6}}
\altaffiltext{$\ast$}{Based on proprietary and
    archival observations with the NASA/ESA Hubble Space Telescope,
    obtained at the Space Telescope Science Institute, which is
    operated by AURA, Inc., under NASA contract NAS 5-26555.}
\altaffiltext{1}{Space Telescope Science Institute, 3700 San Martin
  Drive, Baltimore, 21218, MD, USA}
\altaffiltext{2}{INAF -- Osservatorio Astronomico di Padova, vicolo
  dell'Osservatorio 5, 35122 Padova, Italy}
\altaffiltext{3}{Dipartimento di Fisica e Astronomia `Galileo
  Galilei', Universit\`{a} di Padova, Vicolo dell'Osservatorio 3,
  35122 Padova, Italy}
\altaffiltext{4}{Research School of Astronomy \& Astrophysics,
  Australian National University, Mt Stromlo Observatory, via Cotter
  Rd, Weston, ACT 2611, Australia}
\altaffiltext{5}{ Department of Physics and Astronomy, The Johns
  Hopkins University, 3400 North Charles Street, Baltimore, 21218, MD,
  USA}
\altaffiltext{6}{NASA Goddard Space Flight Center, Exploration of the
  Universe Division, Code 667, Greenbelt MD 20771, USA}

\email{bellini@stsci.edu}

\begin{abstract} 
  We have imaged with \textit{HST}'s WFC3/UVIS the central
  $2\farcm7\times 2\farcm7$ region of the giant elliptical galaxy
  M~87, using the ultraviolet filter F275W. In combination with
  archival ACS/WFC data taken through the F606W and F814W filters,
  covering the same field, we have constructed integrated-light
  UV$-$optical colors and magnitudes for 1460 objects, most of which
  are believed to be globular clusters belonging to M~87. The purpose
  was to ascertain whether the multiple-populations syndrome,
  ubiquitous among Galactic globular clusters (GCs), exists also among
  the M~87 family of clusters.  To achieve this goal, we sought those
  GCs with exceptionally blue UV-to-optical colors, because
  helium-enriched sub-populations produce a horizontal-branch
  morphology that is well populated at high effective temperature.
  For comparison, integrated, synthetic UV$-$optical and purely
  optical colors and magnitudes have been constructed for 45 Galactic
  GCs, starting from individual-star photometry obtained with the same
  instruments and the same filters.  We identify a small group of M~87
  clusters exhibiting a radial UV$-$optical color gradient,
  representing our best candidate GCs hosting multiple populations
  with extreme helium content. We also find that the central spatial
  distribution of the bluer GCs is flattened in a direction parallel
  to the jet, while the distribution of redder GCs is more spherical.
  We release to the astronomical community our photometric catalog in
  F275W, F606W and F814W bands and the high-quality image stacks in
  the same bands.

\end{abstract}


\maketitle

\section{Introduction}
\label{sec:intro}

In recent years the discovery and characterization of multiple stellar
populations in globular clusters (GCs) has given a new impetus to GC
studies, opening entirely new scenarios for their formation. First
came the discovery that the main sequence (MS) of $\omega$~Cen splits
into two, distinct parallel sequences (Bedin et al. 2004). Subsequent
spectroscopic analysis showed that, instead of a lower metal
abundance, the blue sequence has a \textit{higher} metallicity
compared to the red sequence, inescapably demanding that blue MS stars
must be greatly enriched in helium (Piotto et al. 2005: see also
Norris 2004). This is now quantified in $Y=0.39$ (King et al.\ 2012),
compared to the metal-poor population that is presumed to have
near-primordial helium abundance ($Y\sim0.24$).

Since then, evidence for multiple populations in GCs has rapidly
accumulated, mainly thanks to the exquisite photometric accuracy of
the Wide-Field Channel of the Advanced Camera for Surveys
(ACS/WFC). NGC~2808 was found to have three main sequences (Piotto et
al. 2007), in spite of showing no dispersion in the abundance of
iron-peak elements, again demanding that each MS have a different
helium content, up to $Y\sim0.40$. In addition, multiple sub-giant
branches were found in several clusters (Milone et al. 2008; Piotto
2009; Piotto et al. 2012; Bellini et al. 2013), as well as a multiple
main sequences (Milone et al. 2010, 2012ab, 2015; Bellini et al. 2013:
Piotto et al. 2015).

Besides shaking the old paradigm of GCs as simple stellar populations,
these findings raise three fundamental questions:\ 1) Is the
multiple-population phenomenon a general property of GCs, no matter in
which galaxy they are hosted today?  2) How were such multiple
populations generated?  3) What is the origin of the helium
enrichment?  The answers to these questions have deep implications for
our understanding of the formation of GCs, i.e., of objects that are
still playing a pivotal role in so many astrophysical areas and
especially for the early star formation in the Universe. Indeed, GC
ages place their formation tantalizingly close to the epoch of
re-ionization, to speculate they (or their precursors) may have had a
part in it.  Meanwhile, the discovery of multiple populations with
large helium differences has shed new light on one long-standing
mystery:\ the puzzling morphology of the horizontal branch (HB) and
its poor correlation with cluster metallicity, an issue that, after
van den Bergh (1967), is traditionally referred to as the
\textit{Second Parameter} problem. Indeed, the helium abundance has
strong control over the range of effective temperatures covered by
stars during their HB phase, with higher helium leading to higher
effective temperatures for HB stars, hence brighter cluster
UV luminosities. Thus, high-helium multiple main sequences are
associated with multimodal distributions of stars on the HB, with
prototypical examples being offered by $\omega$~Cen and
NGC~2808. Multimodal HB distributions offered in fact the first hint
for high-helium sub-populations in NGC~2808 (D'Antona \& Caloi 2004),
which was then confirmed by the \textit{HST} discovery of multiple
main sequences in the same cluster (Piotto et al. 2007).

Most clusters with multiple populations are chemically homogeneous in
iron and other heavy elements, which implies that helium enrichment
and CNO processing have proceeded without being contaminated by
supernova products. This excludes in most cases massive stars
($M\gtrsim 10$ M$_\odot$) from having been involved in the chemical
enrichment of secondary populations, in either their fast rotating
(e.g., Krause et al. 2013) or binary (de Mink et al. 2009) versions
(e.g., Renzini 2008, 2013, for critical reviews). This leaves
intermediate-mass AGB stars as possible sources of the helium-enriched
and p-capture processed material out of which second generations would
have been formed. However, the AGB scenario is also far from providing
a detailed description of the sequence of events leading to GCs and
their multiple populations as we see them today. Particularly
intriguing is the mass budget requirement in this scenario, as the
first generation delivers only $\sim 5\%$ of its mass as AGB ejecta,
whereas secondary populations in GCs tend to be nearly as massive as
--or even more massive than-- the first generation.  Thus, for the
first generation to produce enough AGB ejecta (to account for the mass
of secondary generations) it should have been at least $\sim 20$ times
more massive than its fraction still bound to the cluster. This is
indeed a lower limit, as it assumes 100\% efficiency in converting AGB
ejecta into second-generation stars.  Thus, in this scenario
precursors to present-day GCs should have been much more conspicuous
objects than their surviving remnants.

A mechanism that in principle could alleviate this mass-budget problem
is accretion of AGB ejecta on stars belonging to the same (first)
generation, a process first proposed to account for the composition
anomalies known at that time (D'Antona et al. 1983; Renzini 1983), and
now still entertained in the modern context (Bastian et
al. 2013). However, accretion will inevitably differ from star to
star, hence may produce some spread in chemical composition, but
cannot generate discrete, {\it quantized} sub-populations (Renzini
2013), such as e.g., the seven or so seen in $\omega$ Cen (Bellini et
al. 2010) or the three seen in NGC 2808 (Piotto et
al. 2007). Actually, recent WFC3/UVIS data reveal the presence of at
least five distinct populations in NGC~2808, while M~2 exhibits at
least seven of them (Milone et al. 2015).  This discreteness of
multiple populations is an extremely powerful discriminant for the
various scenarios proposed for the origin of multiple populations,
excluding those which for their very nature are incapable of producing
discrete, distinct populations. For example, in the
fast-rotating-massive-star option (e.g. Krause et al. 2013 and
references therein) stars of secondary populations would form in the
extruding disk of massive stars, with each massive star delivering
secondary stars with a range of abundances. Globally, just a spread of
compositions would result. Moreover, membership into GCs of fast
rotating massive stars would be irrelevant for the production of
secondary stars, which then should be equally abundant in GCs as in
the Galactic halo field, contrary to observations (Renzini
2013). Discreteness of secondary populations therefore demands
discrete star formation events, such as a series of star-formation
bursts out of an interstellar medium (ISM) whose composition is slowly
changing, interleaved by phases during which the mass of the ISM grows
thanks to the ejecta from the first generation.

In any event, the requirement of a 20--100 times more massive
precursor to present day GCs remains valid in the AGB scenario, thus
making GC formation a major event in the early stages of galaxy
evolution.  Such precursors may have been dwarf nucleated galaxies,
$10^7$--$10^8$ M$_\odot$ by mass in which, past the supernova era, the
AGB ejecta stream to the bottom of the potential well in a sort of
{\it cooling flow} to form the secondary populations, such as e.g. in
the models of D'Ercole et al. (2010, 2011). In these models most of
the stars of the parent dwarf are later stripped by tidal
interactions, thus leaving the bare nucleus with its multiple
populations. One pressing question is whether this process of GC
formation requires special environmental circumstances, which may be
present in one (proto)galaxy but not in others. In our own Milky Way
(MW), the multiple-population syndrome is virtually ubiquitous among
GCs (Piotto et al. 2015), but what about other galaxies? And
especially, what about very massive elliptical galaxies with their
thousands of GCs?  It is indeed quite important for our understanding
of GC formation to establish whether the presence of multiple
populations is a generic property of GCs, irrespective of the type and
mass of the parent galaxy, or whether a special environment (parent
galaxy) is required. To answer this question one will have to rely on
integrated light observations, as resolving stars in GCs is not
currently feasible for distances beyond the Local Group.

The existence of helium-enriched sub-populations can be inferred from
the presence of extremely hot HB stars (EHB), which dominate the UV
luminosity of clusters containing them in sufficient numbers. Typical
in this respect are the metal-rich clusters NGC~6388 and NGC~6441 that
belong to the Galactic bulge, where the existence of EHB stars was
first inferred from their blue UV$-$optical colors and then confirmed
by \textit{HST} photometry (Rich et al. 1997; Busso et al. 2007). In
this way, the integrated UV$-$optical colors of GCs were thought
  to offer a unique opportunity to infer the existence of helium-rich
sub-populations in distant, unresolved GCs belonging to galaxies of
various morphological types, which cannot be resolved into individual
stars. In this mood, the bluer the UV$-$optical color at a given
$V-I$ color, the stronger the EHB component, and especially so for the
reddest $V-I$ clusters. A first such experiment was attempted by Sohn
et al. (2006, hereafter S06) and Kaviraj et al. (2007, hereafter K07)
for the giant elliptical M~87, the cD galaxy of the Virgo
cluster. These authors used STIS FUV (covering the 1150--1700
\AA\ range) and NUV (1600--3100 \AA\ range) data in combination with
optical photometry to infer the likely existence of helium-rich
sub-populations in a small group of GCs in M~87. The presence of a
very strong EHB component in these clusters was inferred from
their extremely blue UV$-$optical colors $\rm{FUV}-V$ and $\rm{NUV}-V$
for a given $V-I$. In these studies M~87 GCs appear to be even more
UV-strong than the bluest Galactic GCs with helium-rich populations,
thus suggesting that the multiple-population syndrome is spread also
among GCs in this giant elliptical.

The S06 and K07 studies have opened the tantalizing opportunity to
diagnose the presence of (likely) helium-enriched extreme EHB stars in
very distant globular clusters, and in particular in a giant
elliptical galaxy. Yet, their sample includes only 66 M~87 clusters,
due to the very small area covered by the STIS observations (just
$\sim 2540$ arcsec$^2$). Moreover, for the reference population, they
had to rely on early integrated-light UV observations of Galactic GCs,
namely with the {\it Astronomical Netherlands Satellite} (van Albada
et al. 1981), with different UV passbands and calibrations compared to
the STIS data used for the M~87 GCs.

For these reasons, we submitted a \textit{HST} proposal (GO-12989,
PI:\ A.~Renzini) to observe the central part of M~87 with the
Ultraviolet-VISible channel of the Wide-Field Camera 3 (WFC3/UVIS) in
the F275W band, exploring in a single shot an area over $10$ times
larger than that covered by the STIS observations and reaching
substantially fainter objects. Moreover, for our reference population,
we can also use WFC3/UVIS ultraviolet data for both M~87 and the MW GC
systems, thus avoiding passband differences and calibration issues
when comparing M~87 and MW clusters.  The complementary archival
optical data consists of very deep (50 orbits) archival F606W and
F814W ACS/WFC imaging of the central region of M~87 (GO-10543,
PI:\ E.~A.~Baltz).

Thus, the plan of the project was to construct UV$-$optical
color-magnitude diagrams and a $m_{\rm F275W} - m_{\rm F606W}$ 
  and $m_{\rm F275W} - m_{\rm F814W}$ vs.\ $m_{\rm F606W} - m_{\rm
  F814W}$ two-color diagrams to characterize the UV properties of a
large, representative sample of GCs in M~87.  Our aim was then to
  address the following questions:
\begin{itemize}
\item{Is the presence of helium-enriched GC sub-populations as
  widespread (or more so) in a giant elliptical compared to the Milky
  Way?}
\item{Has environment (present parent galaxy type) affected the GC
  formation process by favoring/disfavoring the formation of multiple
  stellar generations?}
\item{Is the frequency of UV-bright GCs different among the blue/red
  component of the bimodal color (metallicity) distribution of GCs in
  a giant elliptical such as M~87?}
\item{Is there a connection between the UV-bright populations in GCs
  and the UV-upturn in elliptical galaxies?}
\end{itemize}

However, answering these questions proved to be more difficult than we
expected. Indeed, the \textit{HST} F275W data acquired in the meantime
for a large sample of MW GCs (Piotto et al. 2015) show that the above
UV$-$optical two-color plots are not able to provide unambiguous
evidence for the presence of an EHB component. Still, the UV data on
the M87 and MW GCs are used to highlight possible systematic
differences between the two cluster families. UV$-$optical color
gradients within the brightest M~87 clusters are also used to
tentatively find evidence for the presence in them of helium-enriched
EHB stars.

The paper is organized as follows. Sections~2 and 3 describe the
\textit{HST} data reduction and processing. The photometric
measurement of GCs in M~87 is presented in Section~4 while Section~5
deals with the completeness of the sample. Section~6 presents the
color-magnitude and two-color plots for the M~87 globulars, whose
spatial distribution is discussed in Section~7. Section~8 offers an
extensive comparison with the Milky Way GCs and Section~9 investigates
the possibilities of the detection of internal color gradients within
individual GCs. Finally, in Section~10 we present and discuss our
interpretation of the data and our conclusions.

\section{Data sets and Preliminary Reduction}
\label{sec:dataset}

As mentioned earlier, this work is based on the WFC3/UVIS data from
GO-12989 and on the ACS/WFC data from GO-10543 providing
\texttt{\_flt}-type images\footnote[1]{\texttt{\_flt} images are
  produced by the standard \textit{HST} calibration pipeline
  \textit{CALWF3} (for WFC3) or \textit{CALACS} (for ACS).  Images of
  type \texttt{\_flt} are dark- and bias-subtracted and flat-fielded,
  but not resampled (like the \texttt{\_drz}-type images).} of the
core of M~87.  These data sets consist of $20\times 1365\,$s exposures
taken through the WFC3/UVIS F275W filter and $58\times 500\,$s ACS/WFC
F606W exposures plus $226 \times 360\,$s ACS/WFC F814W exposures. All
\texttt{\_flt} images were corrected for charge-transfer efficiency
defects (e.g., Anderson \& Bedin 2010) using either the standard
\textit{HST} calibration pipeline (for ACS), of the official
\texttt{FORTRAN} routine developed for the
WFC3\footnote[2]{\url{http://www.stsci.edu/hst/wfc3/ins\_performance/CTE/}.}.

The initial catalog of sources in the ACS/WFC exposures was obtained
using the publicly-available \texttt{FORTRAN} program
\texttt{img2xym\_WFC.09x10}, which is described in detail in Anderson
\& King (2006).  The program employs empirical ``library'' effective
point-spread functions (ePSFs) that vary spatially across the
detector.  Source positions and fluxes on WFC3/UVIS images were
measured with the software program \texttt{img2xym\_wfc3uv}, derived
from \texttt{img2xym\_WFC.09x10}, which also employs library,
spatially-varying ePSFs.  Both ACS and WFC3 software identify a source
if its central pixel is brighter than the surrounding 8 pixel,
regardless of the actual shape of the source (point-like, diffuse,
etc.).  Source positions in each single-exposure catalog were then
corrected for geometric distortion using the state-of-the-art
solutions provided by Anderson \& King (2006) for ACS/WFC and by
Bellini \& Bedin (2009) and Bellini, Anderson \& Bedin (2011) for
WFC3/UVIS.

Even though M~87 GCs are slightly resolved by \textit{HST} --and
therefore a PSF is generally not an optimal representation of their
light profiles-- ePSF-derived positions for bright GCs are still found
to be accurate to $\sim 0.02$ pixel.

\begin{figure*}[t!]
\centering
\includegraphics[width=11.5cm,angle=90]{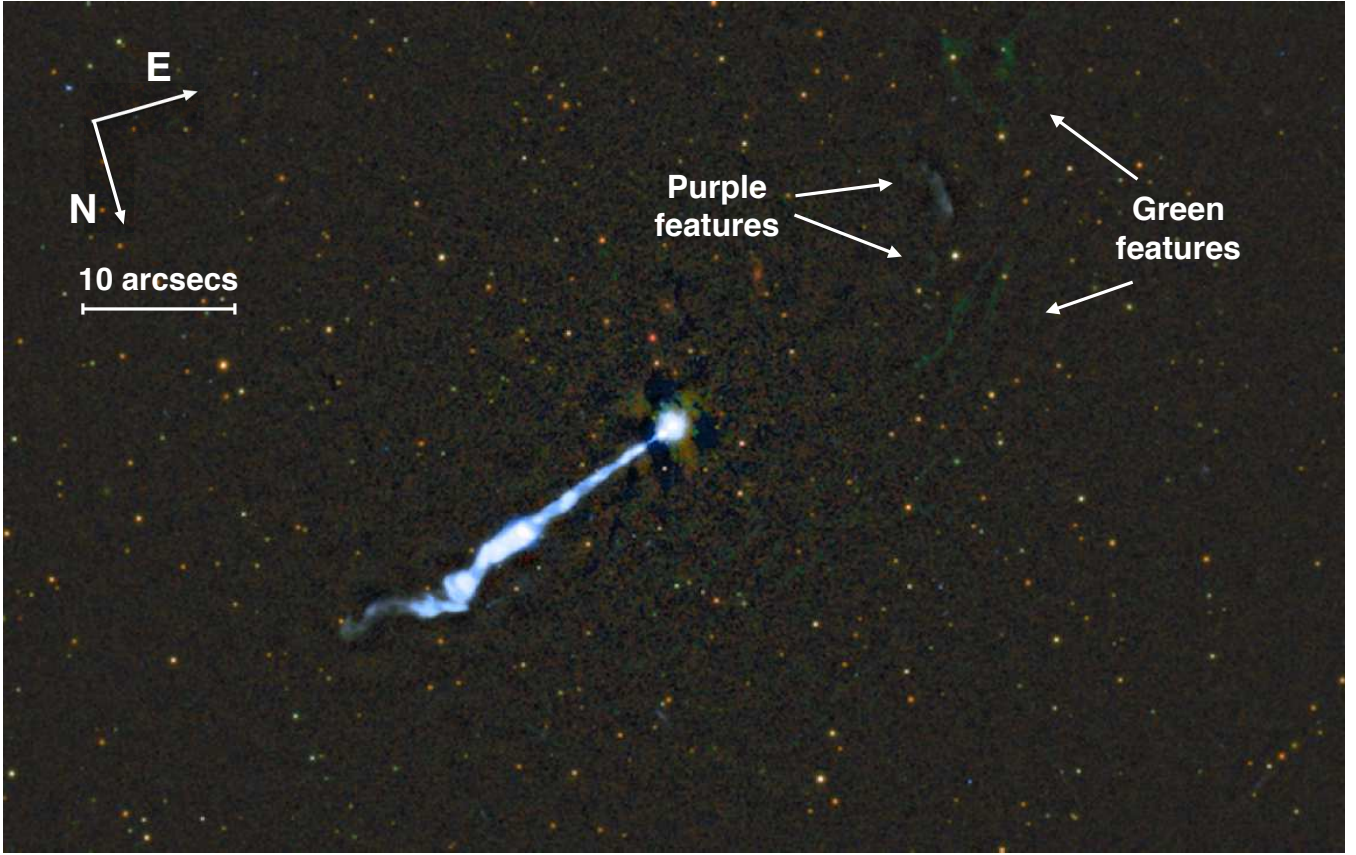}
\caption{The galaxy-subtracted, trichromatic image obtained by
  combining F275W, F606W and F814W stacks.  The color scale has been
  stretched to enhance faint sources and maximize the color difference
  between blue and red globular-cluster candidates.  The vast majority
  of the individual sources are indeed globular clusters in M~87. Some
  counterjet features are also highlighted. These and other features
  are briefly discussed in the dedicated Appendix. Note that this is a
  high resolution image and some of these faint features are better
  seen by enlarging the image on the screen.}
\label{f:gas}
\end{figure*}

\section{Image Stacks}
\label{sec:stacks}

Our final photometry was obtained from high-quality image stacks.
Photometry on the stacks will allow us to obtain a more reliable
measure for fainter sources, and to push down the detection limit by
$\sim$1.6, 2.2 and 3 mag for F275W, F606W and F814W, respectively.

The preliminary single-exposure catalogs have been used only to setup
a common distortion-free reference frame and to define the
linear-transformation parameters that map each pixel of each exposure
onto the image stacks.  We based our reference frame on the F275W
exposures, and we supersampled its pixel size by a factor of 2. This
means that each pixel on the stacks corresponds to 20 mas.  In order
to accommodate all meaningful pixels on the stacks, we shifted the
rough center of M~87 (as defined by the telescope pointing) to
location (5000, 5000) on the master frame. We identified common
sources in each single exposure and used their distortion-corrected
positions to compute general 6-parameter linear transformations to map
each pixel of each exposure on to the master frame.

Although the F275W exposures are each over 22 minutes long, only a few
dozen sources are clearly detected in each of them.  Moreover, due to
the length of these exposures, each F275W image is deeply contaminated
by thousands of cosmic-rays, many of which got included in the
single-exposure catalogs as real objects.  (Cosmic-ray events are far
less abundant in F606W or F814W images, given their shorter exposure
times.)  Special care was dedicated to identifying common sources
between the catalogs to obtain the best linear-transformation
coefficients using only real sources. To this aim, we made use of the
quality-of-fit parameter (Anderson et al.\ 2008) in order properly
select the best-measured objects for the linear transformations.  At
this point, all the necessary pieces of information required by the
stacking algorithms were collected, and we proceeded with the
construction of the image stacks themselves.

Our WFC3/UVIS exposures in F275W were properly dithered with an
image-stacking strategy in mind.  Unfortunately, dithering in the
archival data was minimal.  This means that bad columns, hot/cold
pixels, and other detector defects are often present at the same place
in the field in multiple exposures and cannot simply be
``sigma-clipped'' away.  Therefore, before we could create a stack
that was representative of the astronomical scene, we had to first
identify the warm/hot pixels.  Unfortunately, the DQ
flags\footnote[3]{Each pixel of the \texttt{\_flt} exposures has an
  associated data-quality (DQ) flag that allows the user to
  discriminate between good and bad pixels.}  did not identify all the
warm pixels, so we had to first stack the dithers independently, then
compared each exposure against the stacks made from the other dithers
in order to identify the inconsistent pixels.  We compared each
exposure against the stack made from other dithers and flagged each
time a particular pixel was inconsistent at the $5\sigma $ level with
the prediction from the comparison stack.  We then flagged as bad any
pixel that was flagged in more than half of the exposures.  This
allowed us to identify the bad pixels and construct stacks in F606W
and F814W that were free of artifacts.

Since the goal of this program was to perform aperture photometry on
M~87's globular clusters, we constructed the stacks in such a way as
to carefully preserve flux.  We mapped the four corners of each pixel
in each exposure into the reference frame and determined the geometric
overlap between each input pixel and each output pixel (in the stacked
frame).  These weights were used to construct the output image.  Bad
pixels were naturally given zero weight.  This procedure results in a
small amount of blurring, but this should have a negligible effect on
our photometry given the $\times 2$ oversampling used and the
relatively large apertures we used.

Since each exposure can have a slightly different sky level, on
account of variations in scattered light, we iterated the stacking
procedure.  After each iteration, we resampled the stacked image into
the frame of each individual exposure.  This allowed us to determine
an average sky offset between each exposure and the average of all of
them.  We then adjusted the background level of each exposure by this
difference.  This converged after two iterations and the procedure was
used to generate a stack for the F275W, F606W and F814W exposures.

The cD galaxy in our field introduces a considerable gradient in our
stacks. Since our focus is on the clusters, we decided to iteratively
remove this gradient to make background subtraction trivial. We
parametrized the background by its value at an array of $106\times
106$ elements across the $10\,500 \times 10\,500$-pixel frame. In
other words, we had one array value every $100\times 100$ pixels. We
iteratively solved for the array that reduced the background to zero
everywhere, using a Zeno factor of 0.5 and 16 iterations. The
background was smooth enough to use linear interpolation everywhere.

Figure~\ref{f:gas} shows the central region of M~87 on our
galaxy-light-subtracted trichromatic stack.  The vast majority of near
point-like sources in the figure are M~87 GCs and the color scale has
been stretched to emphasize the color difference between blue and red
GC candidates. The granularity of the background in close proximity of
the galactic center is mostly caused by background noise, but far from
the center the noise is low enough to clearly reveal
surface-brightness variations in the light profile of M~87.  The
famous jet is prominent in this figure and various counterjet features
are also apparent.  The study of the jet and related features goes
beyond the scope of the present paper, but we briefly discuss them in
the Appendix.

\begin{figure*}[ht!]
\centering
\includegraphics[height=8.7cm]{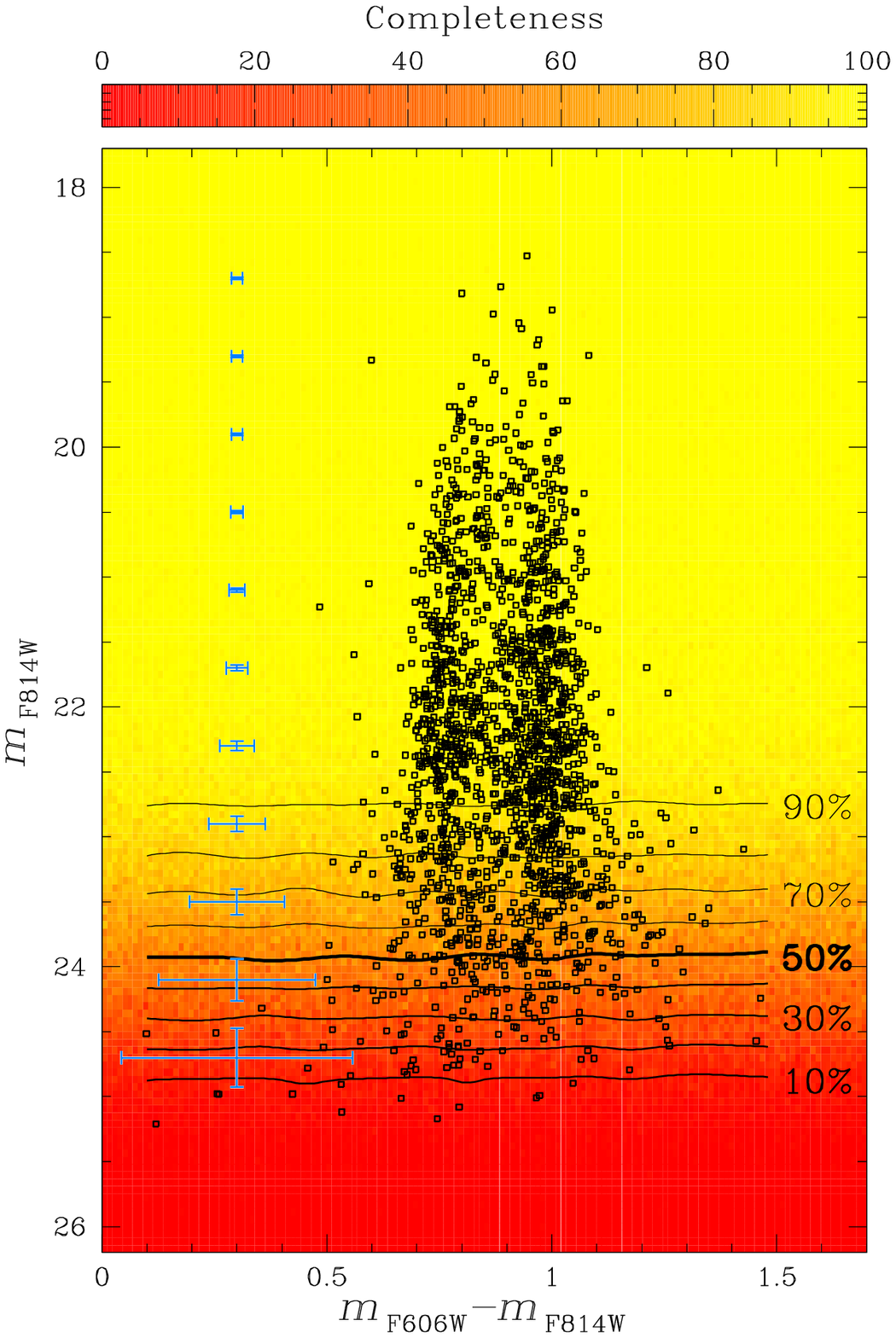}
\includegraphics[height=8.7cm]{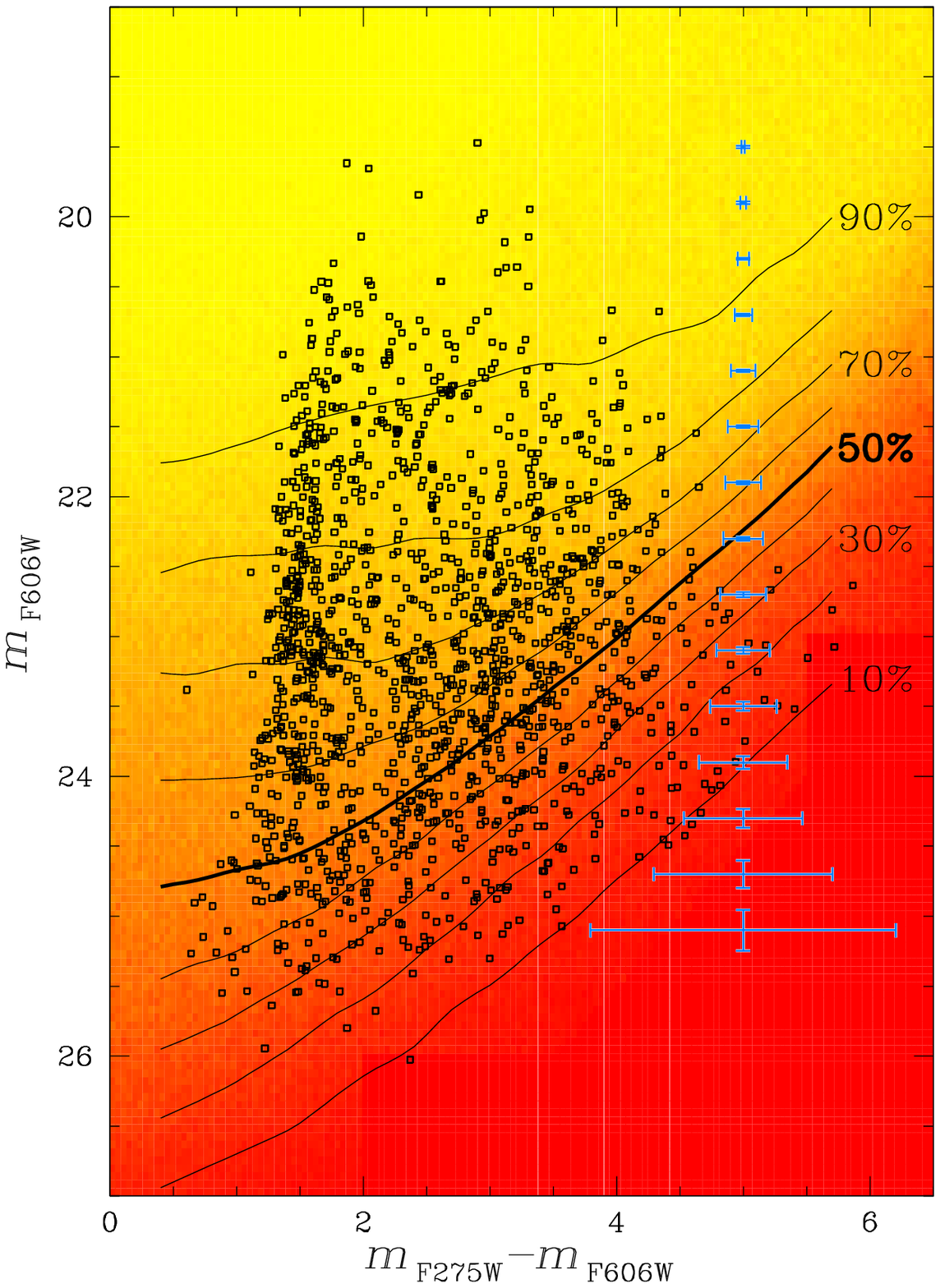}
\includegraphics[height=8.7cm]{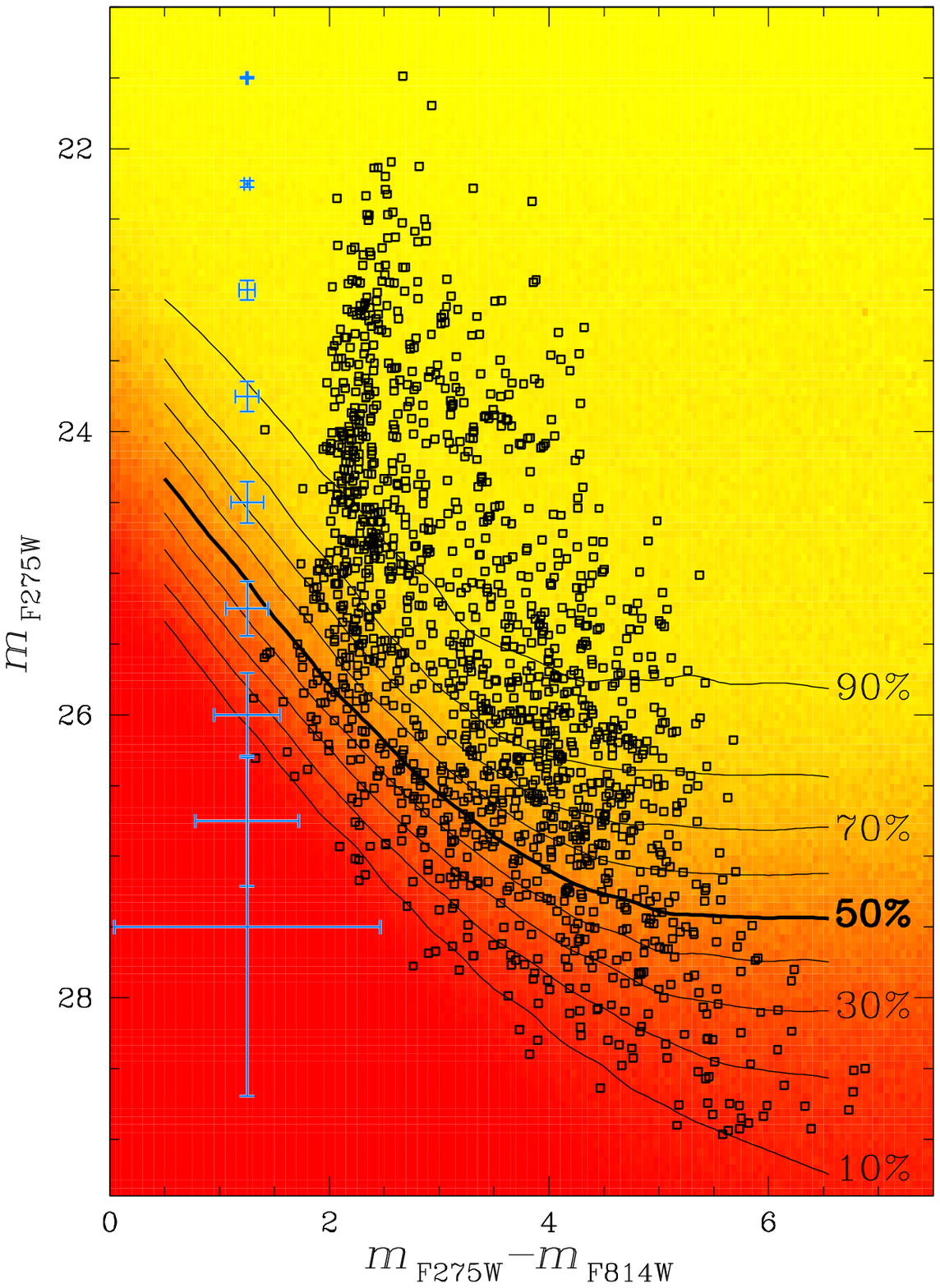}
\caption{The three color-magnitude diagrams for the individual sources
  in the M~87 field. The background is color-coded from red (0\%) to
  yellow (100\%) according to the measured completeness
  level. Photometric error bars, also inferred from the completeness
  analysis, are shown in light blue. Completeness percentiles are in
  black.}
\label{f:complecmd}
\end{figure*}

\begin{figure}[hb!]
\includegraphics[width=\columnwidth]{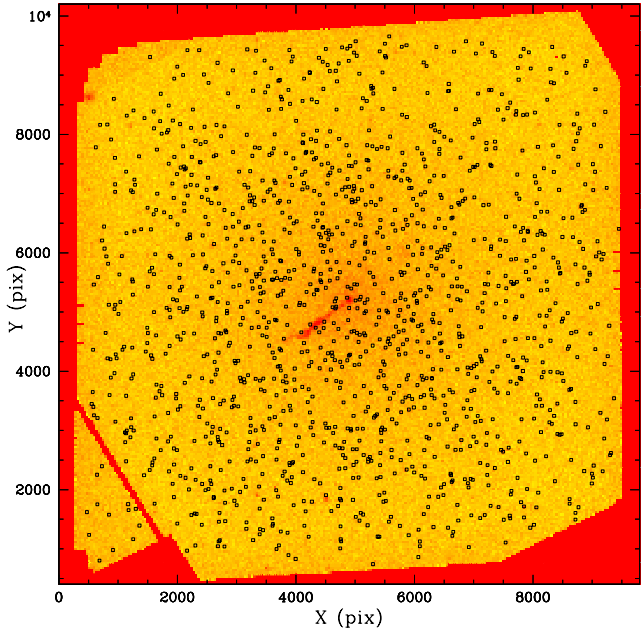}
\caption{The spatial completeness map over the entire field of view,
  color coded as in the previous figure, for the F814W filter. GCs
  measured in all three bands are in black. The completeness level
  decreases toward the galaxy center, and in particular around the
  jet, where it averages 30\% in its close proximity.}
\label{f:complexy}
\end{figure}

\section{Photometry of individual sources}
\label{sec: photo}

As already mentioned in Section~\ref{sec:dataset}, M~87 GCs are
partially resolved, making PSF-fitting techniques not appropriate for
high-precision photometry. Therefore, we applied standard aperture
photometry to measure the flux of the M~87 GCs.

The finding process was performed on the F814W stack only:\ this stack
is made by the largest number of single exposures, and it should
therefore be the one with the highest signal-to-noise ratio.  We
identified all pixels in each exposure that were local maxima (larger
than their 8 surrounding neighbors) and were more than 2.5 times the
rms of the background noise. We ignored areas of the field where there
was fewer than 3 exposures' depth.

A total of 2155 qualifying peaks were found this way. A visual
inspection of their locations on the F814W stack allowed us to remove
spurious objects (mostly obvious galaxies and stars) from the list,
leaving 1913 GC candidates.  Aperture photometry was then performed at
these locations on the three image stacks, using a circular aperture
with a 7-pixel radius ($0\farcs14$). The local sky was estimated
between 15 and 20 pixels.  The choice of using a 7-pixel aperture is
motivated by the fact that it encircles the most-significant pixels of
typical M~87 GCs (being slightly-resolved sources which typical FWHM
of $\lesssim$5 pixels). A larger aperture would have made our
photometry too sensitive to random background fluctuations, especially
for the faint sources in F275W. For the brightest GCs in M~87, a
larger aperture is more appropriate. A 15-pixel aperture for these
objects is applied and discussed in Section~\ref{ss:mass}.

All of the 1913 objects have positive measured fluxes in F814W and
F606W, and 1460 of them also had positive fluxes in the F275W stack,
making our UV photometry by far the largest available in the
literature for extra-galactic GCs.  Finally, the photometry in the
three bands was calibrated into the Vega-mag flight system following
Sirianni et al.\ (2005), and using the F275W zero point and aperture
correction provided by the STScI
webpage\footnote[4]{\url{http://www.stsci.edu/hst/wfc3/phot\_zp\_lbn}.}.

A visual inspection of the F275W image stack revealed the presence of
a handful of relatively bright UV objects that were not identified in
the F814W stack. These objects are likely transient sources in M~87,
such as novae, or UV-bright background galaxies. Since our aim is to
study M~87 GCs, we will not consider them further in this paper.

\section{Completeness}
\label{sec:comple}

In order to infer the completeness level of our source list, as well
as to estimate photometric errors, we ran extensive artificial GC
tests.  For the simulated GCs we used 2D Gaussians with a full width
at half maximum of 4.5 pixels, which is the average value found for
M~87 GCs.

While it is true that our stacks are made in such a way to preserve
both source positions and fluxes, the actual shape of the GCs in the
stacks is the average of several, uncorrelated PSFs (due to different
telescope-breathing conditions for each exposure).  Thus, a 2D
Gaussian turned out to be a fair representation of these average PSFs
while accounting also for GCs of M~87 not being point-like sources.

Instead of the usual practice of simulating GCs with magnitudes and
colors following the cluster fiducial sequences on the CMDs, we
populated the full color-magnitude diagram (CMD) space occupied by the
real GCs with nearly 17 million sources, and assigned to them a random
position on our stacks. Note that M~87 GCs span over 6 magnitudes in
the $m_{\rm F275W}-m_{\rm F814W}$ color, hence this approach allows us
to better estimate incompleteness and photometric errors across such a
wide color range.  As for the real GCs, the finding process was
performed only on the F814W stack. We added one simulated GC at a
time, and determined whether its central pixel qualifies as a peak. If
so, we measured its flux using a 7-pixel-radius aperture photometry in
all three stacks. Before moving to the measurement of the next
artificial object, we removed the current simulated GC from the
stacks. This way, simulated GCs never interfere with each other, but
only with the real sources of the stacks, as it should be.

We consider a simulated GC to be recovered if its measured position in
the F814W stack was within 0.75 pixel of the input position, and if
its measured photometry was within 0.75 mag of its input photometry in
each of the three bands.  There was no need to impose a position
constraint on the F275W and the F606W recovered magnitudes since,
exactly as we have done for the real GCs, their positions are defined
by the measured centroids on the F814W stack. Finally, the
completeness level was simply obtained as the ratio between the number
of recovered GCs and the number of input GCs.

\begin{figure*}[t!]
\centering
\includegraphics[height=5.9cm]{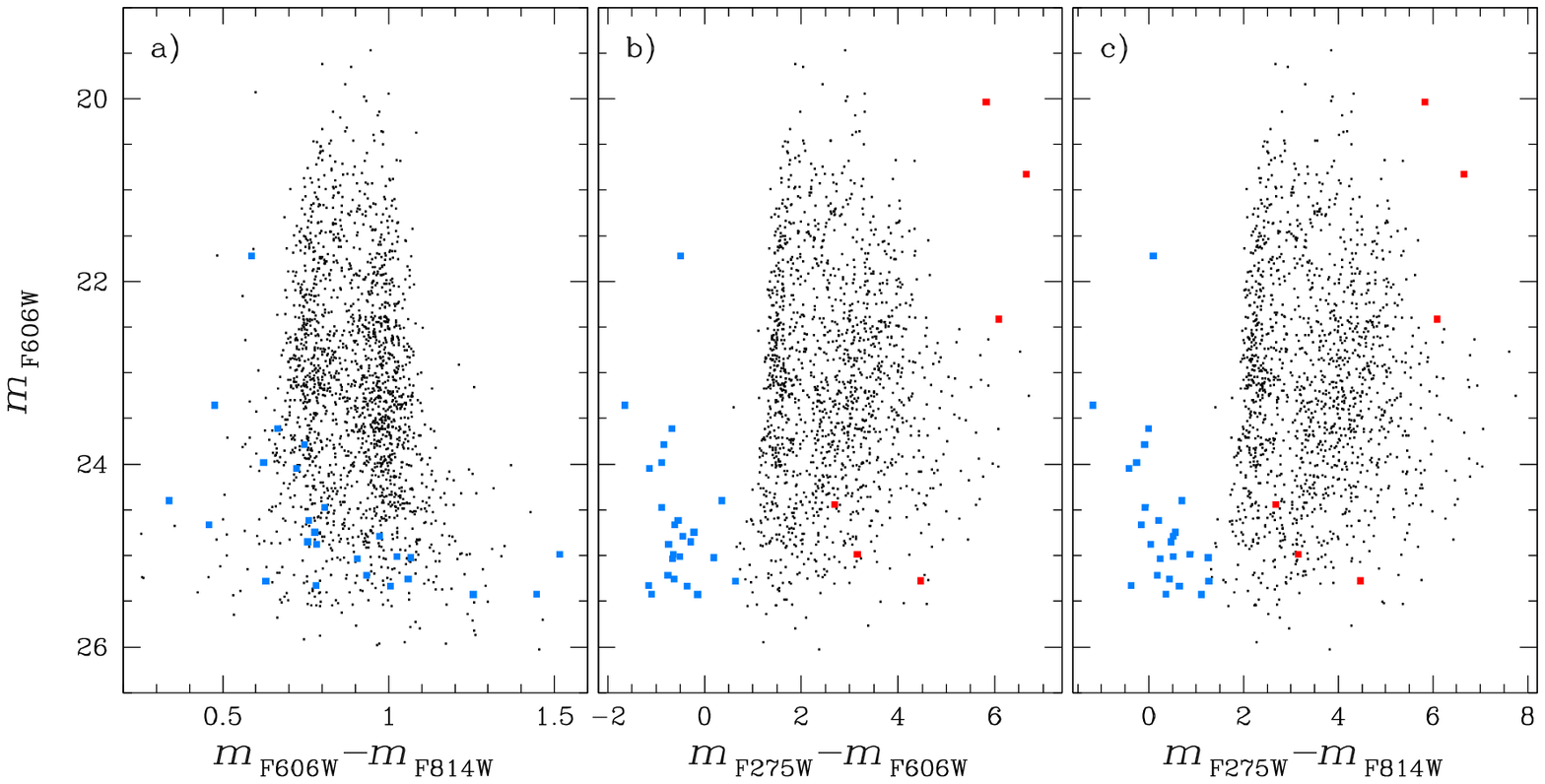}
\includegraphics[height=5.9cm]{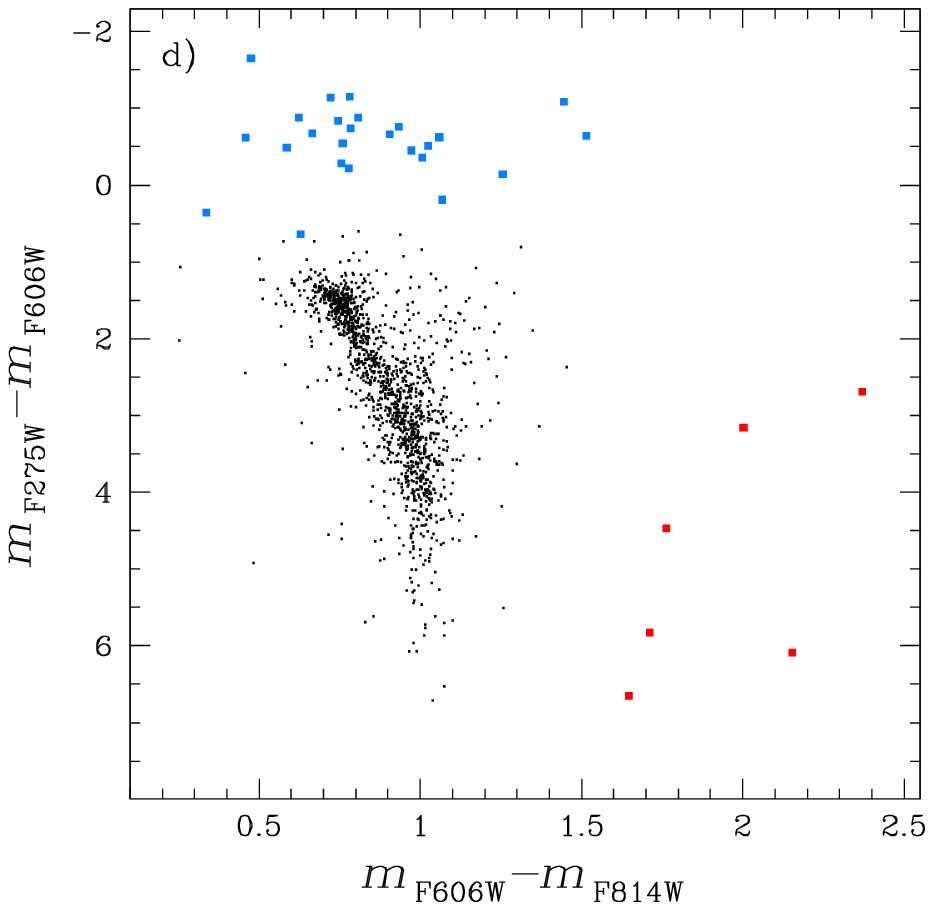}\\
\caption{The optical and UV$-$optical color-magnitude diagrams of the
  sources in the M~87 field (panels a to c) with highlighted in blue
  and red the objects excluded from further discussion, being
  considered star-forming compact galaxies at moderate redshift the
  formers, and compact passive galaxies the latters. Panel (d) shows
  the two-color plot of the same sources.}
\label{f:redblue}
\end{figure*}

To estimate the photometric errors, we computed the 68.27$^{\rm th}$
percentile (rms) of the absolute values of the difference between
input and output magnitudes. These rms values represent our estimate
of the photometric uncertainties.  The results of the artificial GC
tests are summarized in Figures~\ref{f:complecmd} and
\ref{f:complexy}.  The CMDs of real sources (black squares) are shown
Figure~\ref{f:complecmd}, one for each color-filter combination at our
disposal.  In each CMD we divided the simulated GCs in thousands of
small regions (each containing hundreds of input sources), and we
determined the completeness level for each of them. Each region is
then color-coded according to its measured completeness. The color
ranges from red (0\%) to yellow (100\%). A color bar on top of the
first panel illustrates the adopted color-coding, which is kept the
same in all panels. Nine contour lines indicate the loci of average
10\% to 90\% completeness level, with the 50\% contour line being
heavier than the others, for clarity. The contour lines in the optical
CMD are almost horizontal, meaning that, within the investigated color
range, the completeness level of sources measured in the F814W and
F606W stacks is similar, with the F606W stack actually going a little
deeper than that of F814W one. This is not the case for the UV CMDs,
where the completeness level of the sources measured in F275W drops
appreciably towards the red part of the CMDs.

The CMD in which the incompleteness effects are most evident is the
$m_{\rm F275W}-m_{\rm F606W}$ one, where sources have to be found in
all the three bands.  The typical size of the photometric errors are
also shown, in light blue and for sources of average color, in each
CMD. Larger photometric errors are found, as expected, for F275W
measurements.  We will discuss these CMDs in detail in
Section~\ref{sec:cmd}.

Finally, Figure~\ref{f:complexy} shows the spatial F814W completeness
map indicating which parts of M~87 suffer from higher incompleteness
levels regardless of the magnitude level. We adopted the same
color-coding as in Figure~\ref{f:complecmd}. As expected, the
completeness drops towards the center due to the higher background
noise.  The completeness level is also low around the jet, the
location of which can be glimpsed even in this completeness map. The
outer edges of this panel are not mapped by the minimum number of 3
exposures, and no artificial GCs were added there. A red color code
(0\% completeness) is applied to these marginal regions as well.

\section{Color-Magnitude and Two-Color Diagrams}
\label{sec:cmd}

Figure~\ref{f:redblue} shows the CMDs of M~87's GCs using the three
available color combinations (panels a, b and c), as well as the
combined two-color diagram (panel d). The well-known color bimodality
of M~87 GCs (e.g., Gebhardt \& Kissler-Patig 1999) is clearly visible
in all the CMDs. While the two GC sequences have a similar spread in
$m_{\rm F606W} - m_{\rm F814W}$, the blue sequence is significantly
narrower and more defined than the red one in $m_{\rm F275W} - m_{\rm
  F606W}$ and $m_{\rm F275W} - m_{\rm F814W}$. This effect is related
to the increased photometric errors in F275W for the red
GCs\footnote[5]{This effect is far better seen in Figure~5,
    rather than Figure~4, given the more favorable scale and aspect
    ratio.}:\ the color separation between the two GC sequences in
the UV CMDs is $\sim 2$ magnitudes, meaning that, at a given $m_{\rm
  F606W}$ luminosity, red clusters are about 6 to 7 times fainter in
F275W --on average-- than their blue counterparts.

Six detected objects fall outside the red limit of panel (a), being
redder than $m_{\rm F606W} - m_{\rm F814W}=1.6$ (color-coded in red in
the other panels of Figure~\ref{f:redblue}).  They also occupy a
well-defined region in the two-color diagram of panel (d).  A deeper
inspection of their images on the stacks suggests they are compact
passive galaxies at high redshifts and we will not consider them any
further in this paper.

A quick look at panels (b) and (c) of Figure~\ref{f:redblue} also
reveals the presence of a group of 25 extremely-blue objects, and we
color-coded them in blue in all the panels of the figure.  It is worth
noting that these very blue objects seem randomly placed in the
optical CMD, and only the UV photometry allows us to single them
out. These very blue objects occupy a well defined, isolated region in
the two-color diagram shown in panel (d). Most of these blue objects
are faint and appear quite compact in the trichromatic stack:\ we
consider them to be small star-forming galaxies at moderately high
redshift ($z\lesssim 2$), as they are not F275W {\it dropouts}.
Hereafter, we will remove also these 25 objects from our analysis and
therefore we are now left with 1882 likely M~87 GCs, with 1429 of them
having also the F275W photometry.

\begin{figure*}[t!]
\centering
\includegraphics[width=16cm]{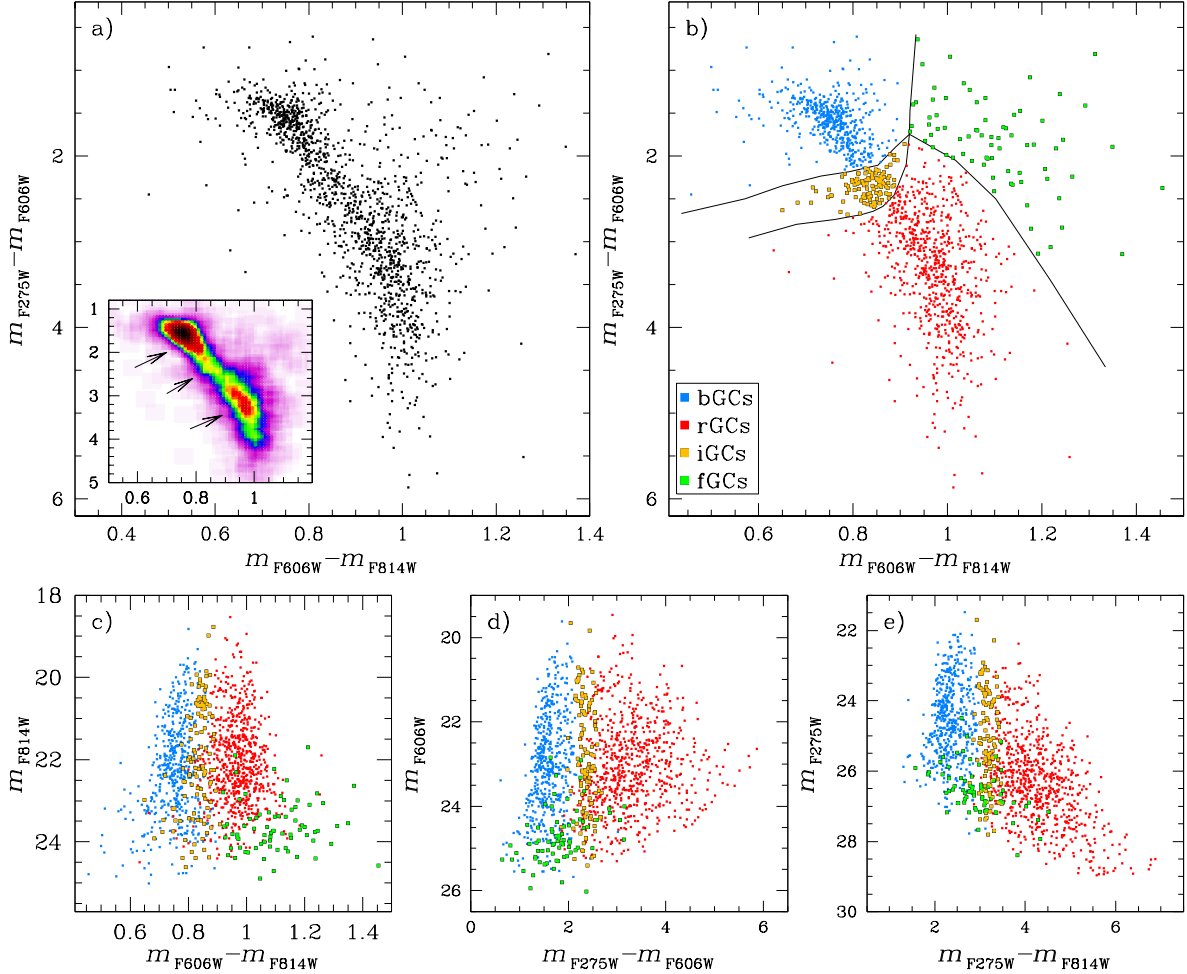}
\caption{Panel (a) shows the two-color diagrams of M~87 GCs. Blue GCs
  clump at (0.75, 1.5) in this plane, while rGCs are more spread and
  stretch across increasing $m_{\rm F275W} - m_{\rm F606W}$ values
  while keeping a roughly constant optical color: an effect driven by
  the increasing F275W errors. An intermediate clump of GCs, at about
  (0.85, 2.5), seems to define an intermediate group of GCs
(although their significance is statistically marginal).
The inset shows a Hess diagram in which the three clumps are
    marked with arrows.  An extended tail of GCs is present in the
  top-right quarter of the panel. Panel (b) is a replica of the
  previous panel, in which we show the regions used to arbitrarily
  divide M~87 GCs in 4 groups. The bottom panels (c), (d) and (e) show
  the CMDs of the four groups, color-coded as in panel (b).}
\label{f:4pops}
\end{figure*}

Let us now turn our attention on the two-color diagram; a zoomed-in
view of it can be seen in panel (a) of Figure~\ref{f:4pops}.  The
inset shows a Hess diagram in which two main clumps are clearly
visible.  There is also a hint for a third, less prominent clump in
between the main two ones, located approximately at (0.85, 2.5). The
three clumps are marked with arrows.  Panel (a) is replicated in panel
(b) where we decided to arbitrarily split the sample into four groups
according to their location in this plot. The {\it blue} GC component
(508 objects, hereafter bGC) occupies a well-defined and compact
region in the top-left corner of the panel, while the {\it red}
component (731 objects, hereafter rGC) is more spread and stretches
across increasing $m_{\rm F275W} - m_{\rm F606W}$ values while keeping
a roughly constant optical color. The {\it intermediate} objects (117
by number, hereafter iGC) that seems to stand out in the Hess diagram
are colored in yellow and appear to be offset in $m_{\rm F606W} -
m_{\rm F814W}$ color with respect to the red component (indeed, they
seem to follow the blue-component trend rather than the red-component
trend), and are separated by a less-populated part of the Hess diagram
at $(m_{\rm F275W} - m_{\rm F606W})\simeq 2.2$ from the blue component
itself. Finally, a small number of objects, colored in green, occupy
the top-right region of the panel (73 objects, hereafter fGC). We
stress that the definition of the intermediate GC component is
arbitrary, and only marginally supported by the relative clump in the
two-color diagram.  Still, we will keep this four-group classification
through the paper, and see whether the analyses presented in the next
sections are able to support (or disprove) the presence of iGCs as a
truly distinct group of clusters in M~87.

Turning now our attention to the lower three panels, we can notice
that while bGCs, iGCs and rGCs are always to the left, middle and
right side of these CMDs, respectively, fGCs are redder --on average--
than rGCs in the $m_{\rm F606W} - m_{\rm F814W}$ CMD, while they
become as blue as the bGCs in $m_{\rm F275W} - m_{\rm F606W}$, and of
somewhat intermediate color between bGCs and iGCs (but definitely
bluer that rGCs) in $m_{\rm F275W} - m_{\rm F814W}$. Hence they {\it
  flip} in color.  If the fGCs had been to be among the brightest GCs,
we would have immediately found the best candidates for the
helium-enhanced, multiple-population GCs, as the most prominent EHB
extensions are found among the most massive GCs in the Milky Way.
Alas, these fGCs are too faint to be the M~87 equivalent of, e.g.,
$\omega$~Cen or NGC~2808 (more in Section~\ref{sec:mw}). We looked at
each of them on the trichromatic stack, and their shapes appear to be
essentially indistinguishable from those of the surrounding GCs of the
same optical brightness level. Their average FWHM is also the same as
that of the other clusters. Moreover, they are also rather isolated,
so that light contamination from brighter neighbors is excluded.

Because fGCs are faint in all bands, the reader can argue that we are
just trying to over-interpret what are just simply photometric-error
effects. While it is true that the larger photometric error of fGCs
can in principle explain their larger spread in the CMDs, they are
never really randomly scattered among the other groups, but are always
clumped in specific regions in each CMD. Nevertheless, most of these
objects are so dim in $m_{\rm F275W}$ that, at the faint limit, their
total flux amounts to only 10 electrons above the sky background in
the F275W stack (sky noise of $\sim 1.3$ electrons).  The fGCs could
be just faint rGCs with large errors in the F275W flux. An intriguing
possibility is that the fGCs are diffuse star clusters, such as those
discovered by Larsen \& Brodie (2000) around two lenticular galaxies,
and already found in the Virgo cluster (Peng et
al. 2006)\footnote[6]{Note that Peng et al. (2006) do not mention M~87
  as harboring a significant number of diffuse star clusters, but
  their observations are much shallower than ours.}.  However, having
measured the size of all clusters with $23<m_{\rm F814W}<24$ we found
no difference between fGCs and the other clusters. Another, perhaps
more likely possibility, is that they are not GCs at all, but compact
emission-line galaxies at moderate redshifts.  Nothing really
conclusive can be said about fGCs until deeper F275W photometry
becomes available.

\begin{figure}[t!]
\centering
\includegraphics[width=\columnwidth]{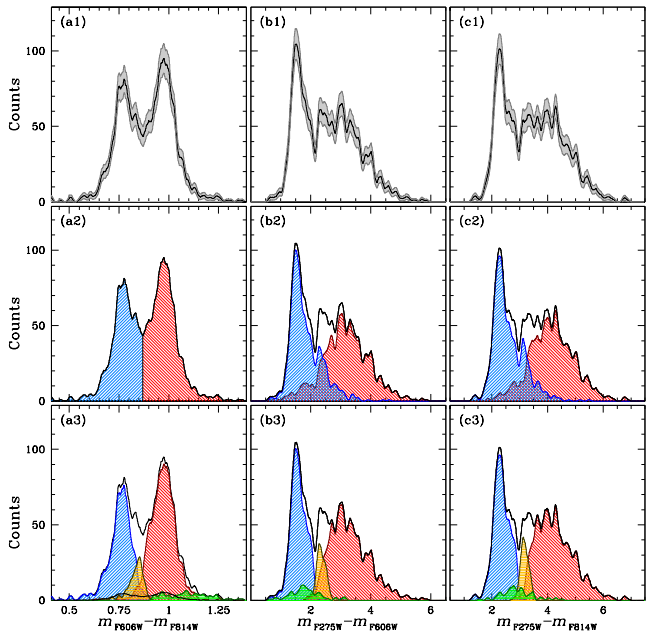}
\caption{Top panels: Histograms of the GC color distributions, with
  Poisson-based confidence regions (in grey). The same histograms (in
  black) are divided into the classical blue/red GC component in panel
  (a2), and color-coded accordingly in (b2) and (c2). In the bottom
  panels we employed the 4-group subdivision introduced in
  Fig.~\ref{f:4pops}. See the text for details.}
\label{f:hist}
\end{figure}

The top panels of Figure~\ref{f:hist} show the histograms of the color
distribution for the three CMDs of Figure~\ref{f:4pops}. We applied a
smoothed na\"{i}ve estimator (Silverman 1986) for the results to be
insensitive to a particular binning starting point.  Two peaks are
clearly distinguishable in panel (a1) referring to the $m_{\rm F606W}
- m_{\rm F814W}$ color distribution, with similar width.  The Poisson
errors are highlighted by shaded regions, and give an assessment of
the quality of the peak detections.  In panels (b1) and (c1) for the
UV$-$optical colors the blue peak is very sharp, while red GCs are
more spread in color and apparently show more than one peak. We
divided the $m_{\rm F606W} - m_{\rm F814W}$ histogram in two halves as
shown in panel (a2), color-coded in blue and red. This represents the
classical bimodal subdivision. We identified the GCs belonging to each
component and color-coded accordingly their histograms in panels (b2)
and (c2). The blue histograms in the middle and right panels both show
a secondary peak at redder colors. The significance of this second
peak is marginal, yet it is present in both the $m_{\rm F275W} -
m_{\rm F606W}$ and $m_{\rm F275W} - m_{\rm F814W}$ colors, which
suggests that it may be a real feature. The lower panels of the figure
adopt the four-group subdivision and the iGC group becomes responsible
for the intermediate peak.

\section{The Globular-Clusters Spatial Distribution}
\label{sec:spatial}

It has long been known that the red GCs of M~87 are more centrally
concentrated than the blue ones (see, e.g., Strader et al.\ 2011), but
previous studies have mostly focused on the clusters' 1D (radial)
surface-density profiles. In this section we investigate the 2D
spatial distribution of M~87 GCs, looking for possible (a)symmetries
in the density maps.

We analyzed the spatial distribution of GCs in two ways:\ first we
simply divided GCs into the two classical red and blue groups, as in
panel (a2) of Figure~\ref{f:hist}; then, we adopted the 4-group
subdvision as defined in panel (b) of Figure~\ref{f:4pops}.  In both
cases, we focus only on GCs above the 50\% completeness limit in the
$m_{\rm F814W}$ vs. $m_{\rm F606W} - m_{\rm F814W}$ CMD, which
corresponds to $m_{\rm F814W} \lesssim 23.9$ mag. By exploring the
spatial distribution of GCs using both the classical and the 4-group
division, we might learn something more about the nature of iGCs and
fGCs, e.g.:\ are iGCs and fGCs two new GC groups, or are they just
subsamples of the classical blue/red components?

\begin{figure*}[t!]
\centering
\includegraphics[width=18cm]{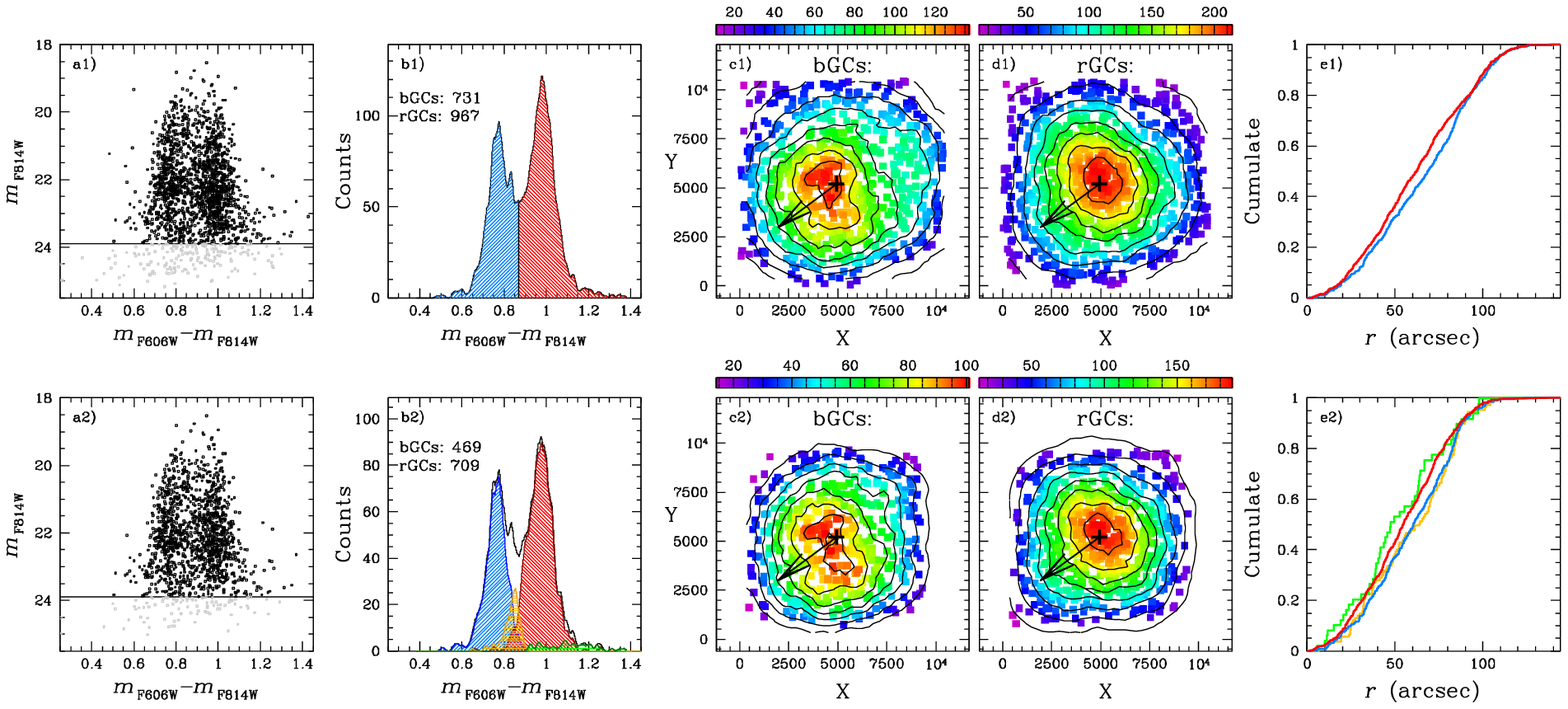}
\caption{Two-dimensional distribution and radial profiles of M~87
  GCs. Top panels refer to the classical blue/red components, while
  bottom panels are for the 4-group subdivision. Panels (a) and (b)
  show the CMD and color histogram of selected stars. Panels (c) and
  (d) show the spatial distribution of bGCs and rGCs in the two cases,
  each with linear-scaled contour levels. The radial profile of each
  GC component is in panels (e).}
\label{f:2dmaps}
\end{figure*}

Let us start from the results of the classical subdivision. Since
blue/red GCs are defined on the $m_{\rm F606W} - m_{\rm F814W}$
without making use of the UV information, to improve the statistical
significance, we included in the analysis all GCs regardless of
whether they also have a $m_{\rm F275W}$ measurement. This way, the
number of GC above the 50\% completeness cut is 1698. Panel (a1) of
Figure~\ref{f:2dmaps} shows the optical CMD for these GCs. The
horizontal line marks the 50\% completeness cut. Rejected GCs are
plotted in grey. Panel (b1) of the figure shows the blue/red GC
subdivision, which is made in the same way as in panel (a2) of
Figure~\ref{f:hist} (the only difference being the 50\% completeness
cut).

To derive the GC density maps we proceed as follows. For each GC in
the two samples we counted the number of surrounding GCs belonging to
the same sample within 1800 pixels ($36^{\prime\prime}$). This radius
was chosen by trial and error as a compromise between the need of a
fine spatial-distribution mapping and an adequate Poisson
statistics. Larger radii would tend to produce very smoothed maps,
washing out most of the intrinsic spatial substructures. On the other
hand, smaller radii tend to produce noisy maps, dominated by
small-number statistics.

Panels (c1) and (d1) of Figure~\ref{f:2dmaps} show the resulting 2D
density maps, for the blue and the red component, respectively.  Each
point in these panels represents the location of selected GCs with
respect to the center of M~87, and it is color-coded according to the
number of surrounding clusters.  Colors range from purple
(low-density) to green (average) to red (high-density).  The center of
M~87 is marked by a black cross and was identified as the brightest
central pixel in the F275W stack (being the center of M~87 heavily
saturated both in F606W and F814W stacks). The direction of the jet is
also shown by a black arrow.

While the rGC density map appears to have a Gaussian-like shape,
roughly centered on M~87, the density map of bGCs appears to be more
asymmetric, and it is flattened in the direction perpendicular to the
jet. So, there is a hint for bGCs being preferentially located about a
{\it galactic plane} roughly perpendicular to the jet.  The 1D radial
profiles of both samples are also shown, in panel (e1), for
completeness.

For the second test we use additional information from F275W
photometry (in order to divide GCs into four groups), thus lowering
the total number of available GCs to 1334 (those above the 50\%
completeness cut).  The optical CMD of selected GCs and their
color-distribution are shown in panels (a2) and (b2) of
Figure~\ref{f:2dmaps}. Unfortunately, the small number of iGCs and
fGCs (107 and 49 objects, respectively) does not allow us to derive
reliable 2D density maps, and we can only compute 1D radial
distributions for them.

As was done for the previous test, for each bGC and rGC we counted the
number of surrounding GCs belonging to the same population within 1800
pixels. The resulting density maps are shown in panels (c2) and (d2)
for bGCs and rGCs, respectively. The available field-of-view (FoV) is
now smaller than that of the first test because of the smaller FoV of
our WFC3/UVIS images with respect to those of the ACS/WFC. For
  both blue and red clusters, in the centermost regions the new
  contours are more flattened than in the case of the previous test
  (and the effect is more evident for the blue GCs), whereas the GC
  distribution in the outer regions does not show a significant
  flattening. This may suggest that iGCs (that were removed in this
  test) may have a more symmetric distribution than blue and red
  GCs\footnote[7]{The small number of iGCs does not allow us to derive
    for them a meaningful density map}.

It is now more evident that in the centermost regions bGCs (and, to a
lesser extend also rGCs) are flattened in a direction nearly parallel
to the jet. In order to better quantify this behavior, we
least-squares fitted with ellipses the contours of panels (c2) and
(d2) to estimate ellipticity and orientation of each contour.  The
angle $\Delta\theta$ between the jet and the minor axis of the
ellipses, and their ellipticities, are reported in Table~1.  The outer
contours suffer from edge effects caused by the square FoV and by the
way we constructed the density maps.  As a consequence, ellipticity
and orientation for the two outermost contours ($r\gtrsim
80^{\prime\prime}$) are not reliable and in Table~1 are reported in
parenthesis. The distribution of both red and blue GCs is more
circular at larger radii, and becomes steadily flatter towards the
center.  The central flattening of blue GCs is about 50\% larger than
that of red GCs. The orientation of the fitted ellipses for the red
GCs is approximately parallel ($\sim -8^{\circ}$) to the jet within
$r\sim 80^{\prime\prime}$.  On the other hand, the orientation of the
ellipses for the blue GCs changes from being parallel to the jet to
being almost perpendicular to it when moving outward.  The agreement
among independent contours, i.e. any contours separated by more than
the region over which the cluster-density was assessed, provides a
sense of the reality of the differences between blue and red GC
subsystems.

Comparing ellipticity and position angle of blue and red clusters with
those of M~87 isophotes (from Liu et al. 2005) we find that the
ellipticities of the GC distributions follow those of the isophotes in
the outer regions of our survey ($\sim 80^{\prime\prime}$), but then
the GC distributions become significantly flattened toward the center
whereas the behavior of the isophotes is opposite, becoming even more
circular within the central $20^{\prime\prime}$. The orientation of
flattening of red and blue GCs is broadly consistent with that of the
outer isophotes of M~87.

The radial profile of all four groups of GCs is shown in panel (e2).
The iGC-supbopulation profile (in yellow in panel e2) is remarkably
similar to that of the bGCs, suggesting that the iGCs might actually
belong to the bGC component, instead of being a separate group.

A special reference needs to be made for the fGC group. They not only
have a unique behavior in the CMDs, but they also appear to be the
most concentrated group of GCs. It is hard to distinguish whether this
indicates their intrinsically different nature or whether they have
larger photometric errors (especially in the F275W band) as on average
they project over a brighter galaxy background.

This flattening near the center of the GC distribution is quite
intriguing indeed, and perhaps even more so that the distribution of
blue GCs is more flattened than that of the red ones. At first, one
would have expected the contrary, with the older, metal-poor component
to be dynamically hotter because of the past merging history of
M~87. However, we cannot exclude that the difference in flattening,
although appreciable, could also arise from a statistical fluctuation
given the relatively small number of objects involved in the
flattening (only 115 bGCs and 200 rGCs in the central $\sim
40^{\prime\prime}$) and the necessary smoothing applied to construct
the density maps.

Given the flattening in the direction of the jet (hence presumably
lying in the equatorial plane of the central supermassive black hole)
this flattening may represent a vestige of an earlier disk morphology
of M~87. However, the galaxy-core dynamics is typical of massive,
non-rotating galaxies which have been built up by merging (see, e.g.,
Figure A1 in Emsellem et al. 2011). Alternatively, rather than a
residual characteristic of the formation of these GCs, the flattening
may have been acquired in the course of the dynamical evolution of the
GC system. For example, one possibility is that clusters in more
radial orbits may have been destroyed by tidal interactions with the
supermassive black hole. We are not in the position to discriminate
between these options and we leave this as an interesting open issue.

\begin{table}[t!]
\begin{center}
\scriptsize{
\begin{tabular}{ccc|ccc}
\multicolumn{6}{c}{\textsc{Table 1}}\\
\multicolumn{6}{c}{\textsc{Ellipses Parameters (inner to outer)}}\\
\hline\hline
\multicolumn{3}{c|}{bGCs}&\multicolumn{3}{c}{rGCs}\\
\hline

Ellipticity$\!\!\!$&$\theta$ (deg)& $\!\!\!$Major axis
($^{\prime\prime}$)&Ellipticity$\!\!\!$&$\theta$ (deg)&
$\!\!\!$Major axis ($^{\prime\prime}$)\\
\hline
0.54   & 13    & 32.6    & 0.34   & -7    & 22.4\\
0.36   & 16    & 50.6    & 0.20   & -6    & 38.9\\
0.20   &  8    & 60.9    & 0.20   & -6    & 51.2\\
0.10   & 30    & 70.5    & 0.12   & -10   & 60.4\\
0.06   & 55    & 79.4    & 0.09   & -8    & 70.4\\
(0.07) & (72)  & (89.5)  & 0.04   & -17   & 79.4\\
(0.07) & (88)  & (101.6) & (0.03) & (-34) & (90.5)\\
       &       &         & (0.01) & (-38) & (103.8)\\
\hline\hline
\end{tabular}}
\end{center}
\end{table}

\section{Comparison with Milky Way Globular Clusters}
\label{sec:mw}

A direct comparison with MW GCs, where the multi-population phenomenon
is now well documented, is mandatory if we are searching for signs of
multiple stellar populations in the GCs of M~87. There are two
\textit{HST} Treasury Programs that made use of the very same
detectors and filters used in this work:\ GO-10775
(PI:\ A.~Sarajedini), which observed 65 MW GCs through ACS/WFC F606W
and F814W filters (Sarajedini et al.\ 2007), and GO-13297
(PI:\ G.~Piotto, Piotto et al.\ 2015), nearly finished and targeting
the cores of 46 out of the GO-10775 65 GCs in three WFC3/UVIS filters,
including F275W. Additionally, images in F275W for other 12 MW GCs
have been acquired with two pilot programs to GO-13297:\ GO-11233,
GO-12311 and GO-12605 (both PI:\ G.~Piotto). Moreover, the core of
$\omega$~Cen was chosen as WFC3/UVIS calibration field, and have F275W
exposures.  At the completion of GO-13297 the total number of clusters
for which WFC3/UVIS F275W and ACS/WFC F606W and F814W data exist will
be 59: about a third of the total number of GCs in the Milky Way.

In order to directly compare MW GCs with those in M~87 we need
to:\ (1) estimate the total light of MW GCs in the three bands; (2)
apply reddening corrections in order to have the same reddening of
M~87 ($E(B-V)=0.04$, obtained by multiplying by 1.35 the quoted value
$E(V-I)=0.03$ of Harris 2009); and (3) bring all MW GCs to the same
distance of M~87 (distance modulus $(m-M)_0=31.0$, again from Harris
2009).

To estimate the integrated total light of the MW GCs we proceeded as
follows.  First of all, we note that all GO-10775 catalogs are
reasonably complete (50\%) down to about 5.5 magnitudes below the
turnoff, and have comprehensive artificial-star tests that allow
proper incompleteness estimates.  A direct sum of the flux of the
cluster members present in these catalogs, within a certain radius, is
therefore a good estimate of the total cluster flux within that
radius. While we will inevitably lose a sizable fraction of stars
fainter than, say the $50\%$ completeness limit (especially near the
cluster center). the dim light of these missed faint sources will only
marginally affect the estimated total light of the clusters and will
have an even smaller impact on the inferred integrated colors.

In fact, using artificial-star tests, we estimated the amount of light
lost from missed stars in F606W and F814W down to the 50\%
completeness level and found it to be on the order of 4\% in both
filters and in nearly all GO-10775 clusters\footnote[8]{Specifically,
  3.9\% (semi-interquartile of 2.0\%) for F606W, and 4.2\%
  (semi-interquartile of 1.8\%) for F814W.}. (This value necessarily
varies spatially, and the light lost in the core due to crowding
effects is higher than in the outer regions. On average, we found a
relative 2\% increase in lost flux within the centermost
$25^{\prime\prime}$ with respect to the $50^{\prime\prime}$ to
$70^{\prime\prime}$ radial interval.) A 4\% difference in total flux
corresponds to about 0.04 magnitudes. The average $m_{\rm
  F606W}-m_{\rm F814W}$ color variation due to incompleteness is then
found to be of the order of only 0.003 magnitudes.  We also computed
the total light of 10 randomly-selected clusters as provided by stars
down to 2, 4 and 6 magnitudes below the turnoff. On average, the total
clusters' light decreased by 0.029, 0.005 and 0.0004 magnitudes for
the 2, 4, and 6 mag cuts in F606W, and by 0.034, 0.007 and 0.0007,
respectively, in F814W.  The difference in the $m_{\rm F606W}-m_{\rm
  F814W}$ color is then 0.004, 0.002 and 0.0003 for the three
magnitude cuts, and therefore negligible.  Given this very small
impact of incompleteness on the total clusters' light and optical
colors, we decided to ignore completeness effects in our calculations.

\begin{figure*}[t!]
\centering
\includegraphics[width=\textwidth]{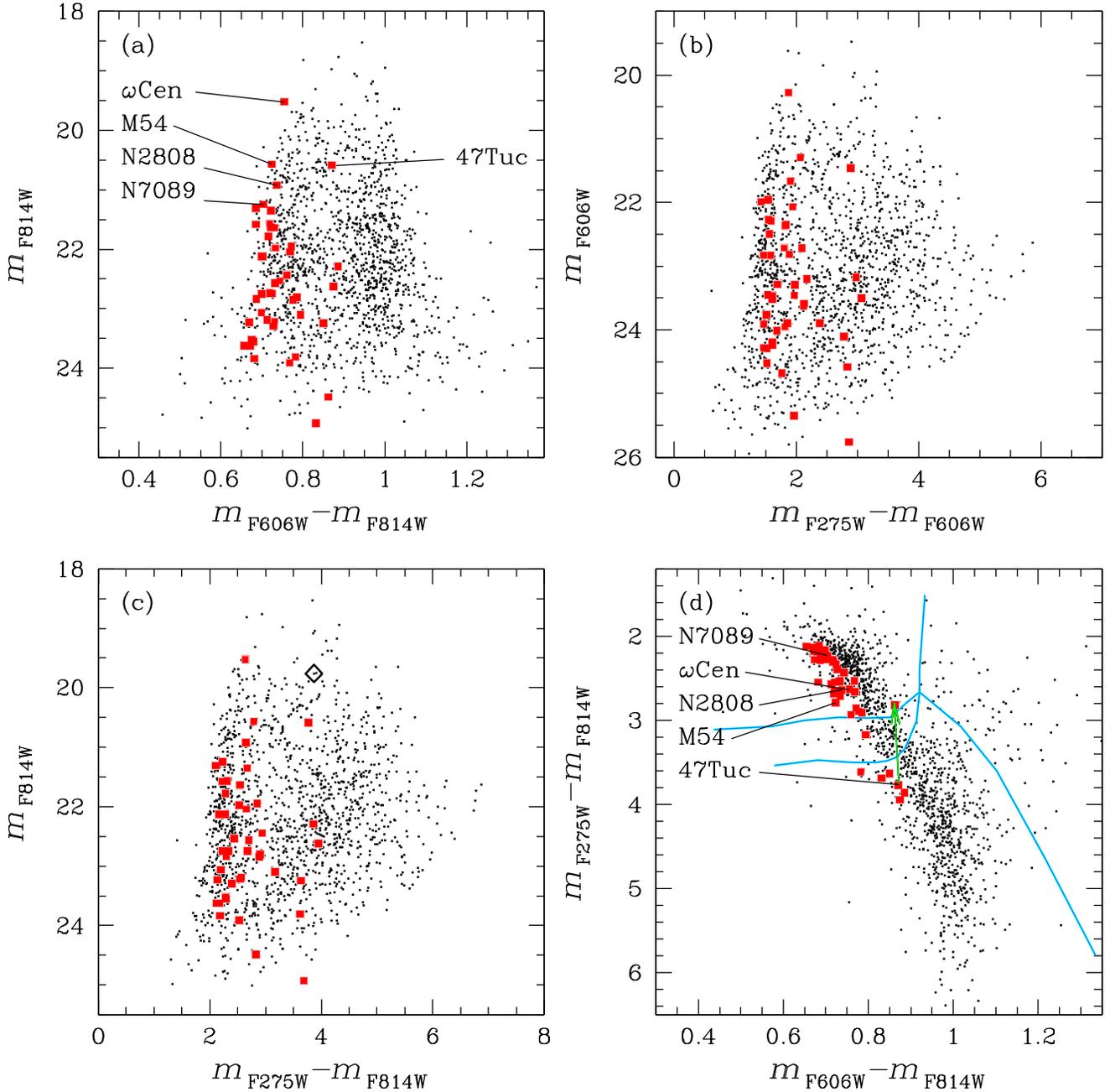}
\caption{Color-magnitude diagrams and two-color plot for the
  integrated light of M~87 GCs (black points) and for 45 MW globulars
  (red squares). Some MW globulars are tagged for clarity.  The
  diamond in panel (c) indicates the cluster with very blue
  FUV$-$F275W color, and is discussed in Section \ref{sec:tony}. The
  blue lines in panel (d) are the same as panel (b) of Fig.~5. The
  green arrow is discussed in Sect.~8.2}
\label{f:mw_piotto}
\end{figure*}

To obtain stellar fluxes in F275W we used the catalogs of GO-10775 as
an input, and fitted F275W library PSFs to each listed stellar
position, so that every source in
the GO-10775 catalogs will also have a F275W measurement if detected
in this band (Soto et al.\ in preparation). We do not yet have
  artificial-star tests for the GO-13297 data. Nevertheless, in F275W
  we can easily detect stars down to 3--4 magnitudes below the
  turnoff, i.e. 10 magnitudes below the HB (in F275W). This typically
  translates into about 2.5 magnitudes below the turnoff in F606W, and
  we already saw that below this cut we lose only a marginal fraction
  of light in the F606W passband.  The missed stars are faint, red MS
  stars that already have little/no contribution to the clusters light
  in optical, and even less in F275W. Unfortunately, this is not the
  case for the two interesting clusters NGC 6388 and NGC 6441, the
  most metal rich clusters with evidence of helium-rich EHB stars
  (e.g., Bellini et al. 2013). These heavily reddened Bulge clusters
  would have required a disproportionate number of \textit{HST} orbits
  to reach an F275W depth below the turnoff similar to that of the
  other clusters. Actually, the data barely reach the turnoff, which
  is a significant contributor to the UV light especially in
  metal-rich clusters, see Figure 10. Therefore, these clusters are
  not included in our present analysis.

For the calculation of the total clusters light we selected only
bona-fide cluster members, on the basis of their proper motions
(obtained as the difference in stellar position between the GO-10775
and GO-13297 catalogs) and on their location on the color-magnitude
diagram. These two selection criteria work very well for all clusters
except for NGC~6715. In this case, it is impossible to separate
cluster members from the core of the Sagittarius dwarf spheroidal
galaxy using proper motions, but at least the CMD selections allowed
us to remove Sagittarius RGB contaminants.

Needless to say, the FoV of GO-13297 catalogs ($\sim 2.7\times2.7$ $
\rm{arcmin}^2$) is smaller than the tidal radius of MW GCs. To convert
our FoV-limited fluxes into the total cluster light, we made use of
the fraction-of-light radii listed in Trager, King \& Djorgovski
(1995). In Their Table~2, the authors report the value of $r_{10}$,
$r_{20}$, $r_{30}$, $r_{40}$, $r_{\rm h}=r_{50}$, which are the radii
within which the 10\% to 50\% of the total cluster light is
enclosed. For each MW GC in our sample, we summed the flux of all the
stars within the largest value $r_{\rm X}$ (where $X$ is one of the
listed fraction-of-light radii) that was still fully enclosed in the
GO-13297 FoV. These $r_{\rm X}$-based fluxes were then accordingly
rescaled to synthesize the total clusters light in each of the three
passbands.  Note that we applied the same scaling factor for UV and
optical magnitudes. If UV light and optical light have different
radial gradients, this might introduce a bias.  Some of our MW
  clusters are not listed in Trager, King \& Djorgovski (1995); we could
  compute integrated clusters lights for only 45 objects.

We adopted the reddening values listed in Harris (1996, 2010 edition),
in order to adjust our integrated magnitudes to the reddening of M~87,
by using the extinction coefficients of Sirianni et al.\ (2005) for
ACS/WFC. To date, no official extinction coefficient is available for
the F275W filter on the \textit{HST} website. We adopted the value
A$_{\rm F275W}=6.14$, obtained with YES, the York Extinction Solver,
using the reddening laws by Cardelli, Clayton \& Mathis (1989).

Finally, we corrected the integrated and (de)reddened magnitudes for
the difference between the MW GCs distance moduli (from Harris 1996,
2010 edition) and that of M~87 ($(m-M)_0=31.0$, Harris 2009).  The
integrated, absolute and de-reddened photometry for MW GCs is listed
in Table~2.

\begin{figure}[t!]
\centering
\includegraphics[width=\columnwidth]{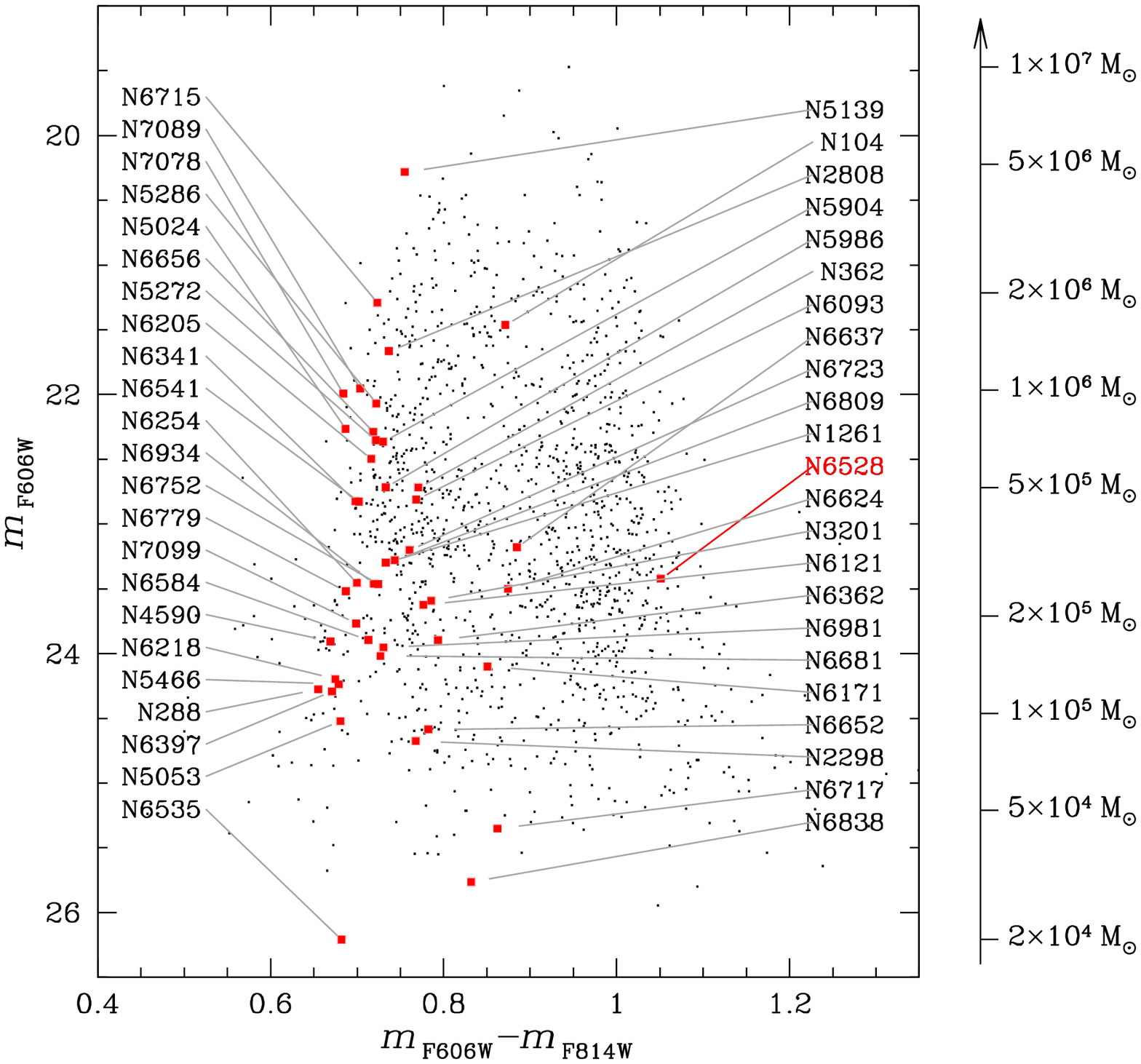}
\caption{The optical color-magnitude diagram of GCs in M~87
  (black points) and in the MW (red squares). In addition, the
  integrated magnitude and color of the Galactic bulge, metal rich
  cluster NGC 6528 is reported, using the astro-photometric catalog of
  Brown et al. (2005).}
\label{f:mw_ata}
\end{figure}

Figure~\ref{f:mw_piotto} shows the three CMDs (panels a, b and c) and
the two-color diagram (panel d) of M~87 (in black) and the 45 MW GCs
(in red), for which we already have the data in all three passbands.
Outstanding MW GCs in the figure are tagged.  Note that none of the 45
MW GCs fall on the bulk of the red sequence of M~87 in $m_{\rm
  F606W}-m_{\rm F814W}$.  The metal-rich GC 47~Tuc ([Fe/H]=$-$0.72,
from Harris 1996, 2010 edition) lies just in the middle between blue
and red M~87 clusters in the optical color, and just makes it on the
red sequence when the color is based on the F275W filter.

Moreover, we note that $\omega$~Cen itself would be among the
brightest GCs in M~87.  Indeed, only a handful of M~87 GCs in our
survey field are brighter than $\omega$~Cen, but not by much more than
1 magnitude in either $m_{\rm F606W}$ or $m_{\rm F814W}$.  It appears
that the most luminous GCs in M~87 are likely siblings of G~1
(Mayall~II) in M~31, with a mass of $\sim 10^7$ M$_\odot$ (Ma et
al.\ 2009).  The integrated color of NGC~6715 (M~54) is somewhat more
uncertain, as this cluster is deeply embedded within the nucleus of
the Sagittarius dwarf spheroidal galaxy, in such a way that there are
a comparable number of cluster and galaxy stars (that are redder, on
average) even in the centermost few arcmin from the cluster center.

From panels (b) and (c) in Figure~\ref{f:mw_piotto} we note that the
bulk of MW and M~87 blue clusters have approximately the same $m_{\rm
  F275W}-m_{\rm F814W}$ color, but in the two-color diagram shown in
panel (d) there is a systematic offset in $m_{\rm F606W}-m_{\rm
  F814W}$ of $\sim 0.06$ mag, also appreciable in panel (a).  Several
sources of error affect our MW GC photometry:\ errors in the
photometric zero-pointing, in the total flux estimate, in the M~87
distance modulus, in the adopted reddening values, and in our estimate
of the total clusters' light.  We can exclude faint star
incompleteness of our MW GC database as the cause of the shift, as
this 0.06 mag difference is much bigger than what would possibly be
expected from such an effect, as estimated above. Another possibility
is that the $m_{\rm F606W}-m_{\rm F814W}$ color of MW GCs may have
been underestimated given that the brightest RGB stars (above the HB
level) tend to be saturated in the GO-10775 images. The photometry of
these RGB stars was corrected for saturation effects, but certainly
the result cannot be as accurate as in the case of unsaturated
stars. Whether this is the reason for the 0.06 mag difference could in
principle be ascertained, but it requires a non trivial effort and we
wish to postpone it to a future paper.  Some of these sources of
systematic errors affect our three photometric measurements in the
same way, but others may indeed generate a systematic color difference
between MW and M~87 clusters.  At this stage, we cannot determine
whether this difference in color of $\sim 0.06$ mag is an intrinsic
property of MW and M~87 GCs or it is due to systematic errors.

\begin{table}[t!]
\begin{center}
\footnotesize{
\begin{tabular}{cccc}
\multicolumn{4}{c}{\textsc{Table 2}}\\
\multicolumn{4}{c}{\textsc{Integrated, reddening-free Absolute Photometry}}\\
\multicolumn{4}{c}{\textsc{of Milky-Way Globular Clusters}}\\
\hline
\hline
GC ID& $m_{\rm F606W}$&  ($m_{\rm F275W}-m_{\rm F606W}$)&  ($m_{\rm F275W}-m_{\rm F814W}$)\\
\hline
NGC~104  & $-$9.65 &  2.76 &  3.60\\ 
NGC~288  & $-$6.84 &  1.33 &  1.95\\ 
NGC~362  & $-$8.40 &  1.95 &  2.68\\ 
NGC~1261 & $-$7.82 &  1.84 &  2.53\\ 
NGC~2298 & $-$6.44 &  1.63 &  2.36\\ 
NGC~2808 & $-$9.45 &  1.77 &  2.47\\ 
NGC~3201 & $-$7.52 &  1.98 &  2.73\\ 
NGC~4590 & $-$7.21 &  1.33 &  1.96\\ 
NGC~5024 & $-$8.85 &  1.41 &  2.06\\ 
NGC~5053 & $-$6.59 &  1.38 &  2.02\\ 
NGC~5139 & $-$10.83&  1.74 &  2.46\\ 
NGC~5272 & $-$8.76 &  1.69 &  2.38\\ 
NGC~5286 & $-$9.04 &  1.81 &  2.49\\ 
NGC~5466 & $-$6.88 &  1.47 &  2.11\\ 
NGC~5904 & $-$8.75 &  1.68 &  2.37\\ 
NGC~5986 & $-$8.40 &  1.67 &  2.36\\ 
NGC~6093 & $-$8.30 &  1.76 &  2.49\\ 
NGC~6121 & $-$7.49 &  1.98 &  2.72\\ 
NGC~6171 & $-$7.01 &  2.65 &  3.46\\ 
NGC~6205 & $-$8.62 &  1.43 &  2.10\\ 
NGC~6218 & $-$6.92 &  1.47 &  2.11\\ 
NGC~6254 & $-$7.66 &  1.39 &  2.05\\ 
NGC~6341 & $-$8.29 &  1.33 &  1.99\\ 
NGC~6362 & $-$7.22 &  2.25 &  3.00\\ 
NGC~6397 & $-$6.82 &  1.38 &  2.01\\ 
NGC~6535 & $-$4.90 &  1.73 &  2.38\\ 
NGC~6541 & $-$8.29 &  1.44 &  2.11\\ 
NGC~6584 & $-$7.22 &  1.72 &  2.39\\ 
NGC~6624 & $-$7.61 &  2.94 &  3.77\\ 
NGC~6637 & $-$7.93 &  2.84 &  3.69\\ 
NGC~6652 & $-$6.53 &  2.70 &  3.44\\ 
NGC~6656 & $-$8.82 &  1.45 &  2.13\\ 
NGC~6681 & $-$7.09 &  1.54 &  2.23\\ 
NGC~6715 & $-$9.82 &  1.93 &  2.62\\ 
NGC~6717 & $-$5.76 &  1.83 &  2.65\\ 
NGC~6723 & $-$7.91 &  2.04 &  2.77\\ 
NGC~6752 & $-$7.65 &  1.48 &  2.16\\ 
NGC~6779 & $-$7.59 &  1.47 &  2.12\\ 
NGC~6809 & $-$7.83 &  1.56 &  2.26\\ 
NGC~6838 & $-$5.35 &  2.73 &  3.52\\ 
NGC~6934 & $-$7.65 &  1.83 &  2.51\\ 
NGC~6981 & $-$7.16 &  1.68 &  2.37\\ 
NGC~7078 & $-$9.12 &  1.29 &  1.94\\ 
NGC~7089 & $-$9.16 &  1.40 &  2.06\\ 
NGC~7099 & $-$7.34 &  1.37 &  2.03\\ 
\hline\hline
\end{tabular}}
\end{center}
\end{table}

In analogy to the result of S06 and K07, we find that M~87 GCs appear
systematically bluer than MW GCs in the color-color plot, at fixed
optical color. However, the blue GCs in both galaxies span the same
range of colors in $m_{\rm F275W} - m_{\rm F814W}$, hence we attribute
the offset to the $m_{\rm F606W} - m_{\rm F814W}$ color, i.e., M~87
GCs do not appear to be {\it more ultraviolet} than MW GCs.  Our data
do not allow us to exclude this offset being due to
systematic/calibration errors. We note that this color shift was more
prominent in S06 and K07 FUV$-$optical color rather than in the
NUV$-$optical color. The STIS NUV passband used by S06 and K07 partly
overlaps with the F275W passband, but the FUV passband is even bluer
than the bluest WFC3/UVIS filter:\ F225W.  Such a FUV passband would
indeed be more sensitive to the bluest HB stars. In Section~10 we
shall return on the comparison with S06 and K07 results.

The fact that the vast majority of the MW GCs in our sample occupy the
blue half of the M~87 GCs in all the three CMDs should not come as a
surprise. The GO-10775 treasury program (and therefore also the
GO-13297 one) did not include heavily-reddened clusters, which are
generally harbored in the Galactic bulge. These clusters are
metal rich and also intrinsically red.

In the Milky Way Bulge there are some extremely-metal-rich clusters
that may lie well within the rGC sequence of M~87. To verify this, we
additionally estimated the integrated light of NGC~6528
([Fe/H]=$-0.11$, $E(B-V)=0.54$, from Harris 1996), an almost
solar-metallicity GC in the Bulge, following the very same procedures
and selection criteria used for the other clusters.  NGC~6528 was
observed through ACS/WFC F606W and F814W as part of the GO-9453
program (PI: T.~M.~Brown, see Brown et al. 2005).  This cluster
(identified and tagged in red in Figure~\ref{f:mw_ata}) lies on the
red side of the red sequence, as one would expect. So, we cannot claim
dramatic systematic differences between color and brightness span of
the MW and M~87 GC sequences.

\begin{figure}[t!]
\centering
\includegraphics[width=\columnwidth,angle=-90]{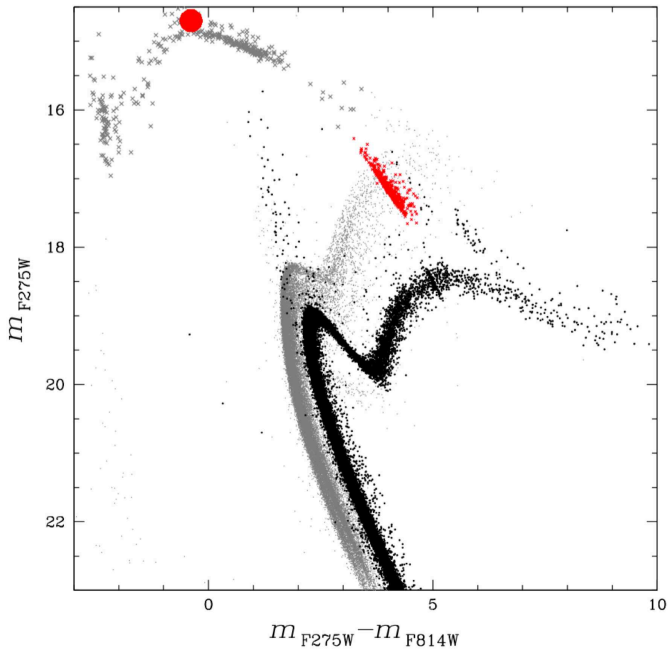}
\caption{The CMDs of $\omega$ Cen and 47 Tuc brought to a common
  distance (grey and black points, respectively) with the HB stars of
  47~Tuc in red.  The large red circle shows where all the red-clump
  stars of 47~Tuc were placed to computed the integrated color
  differences discussed in the text.}
\label{f:47tuc}
\end{figure}

\begin{figure*}[t!]
\centering
\includegraphics[width=15cm]{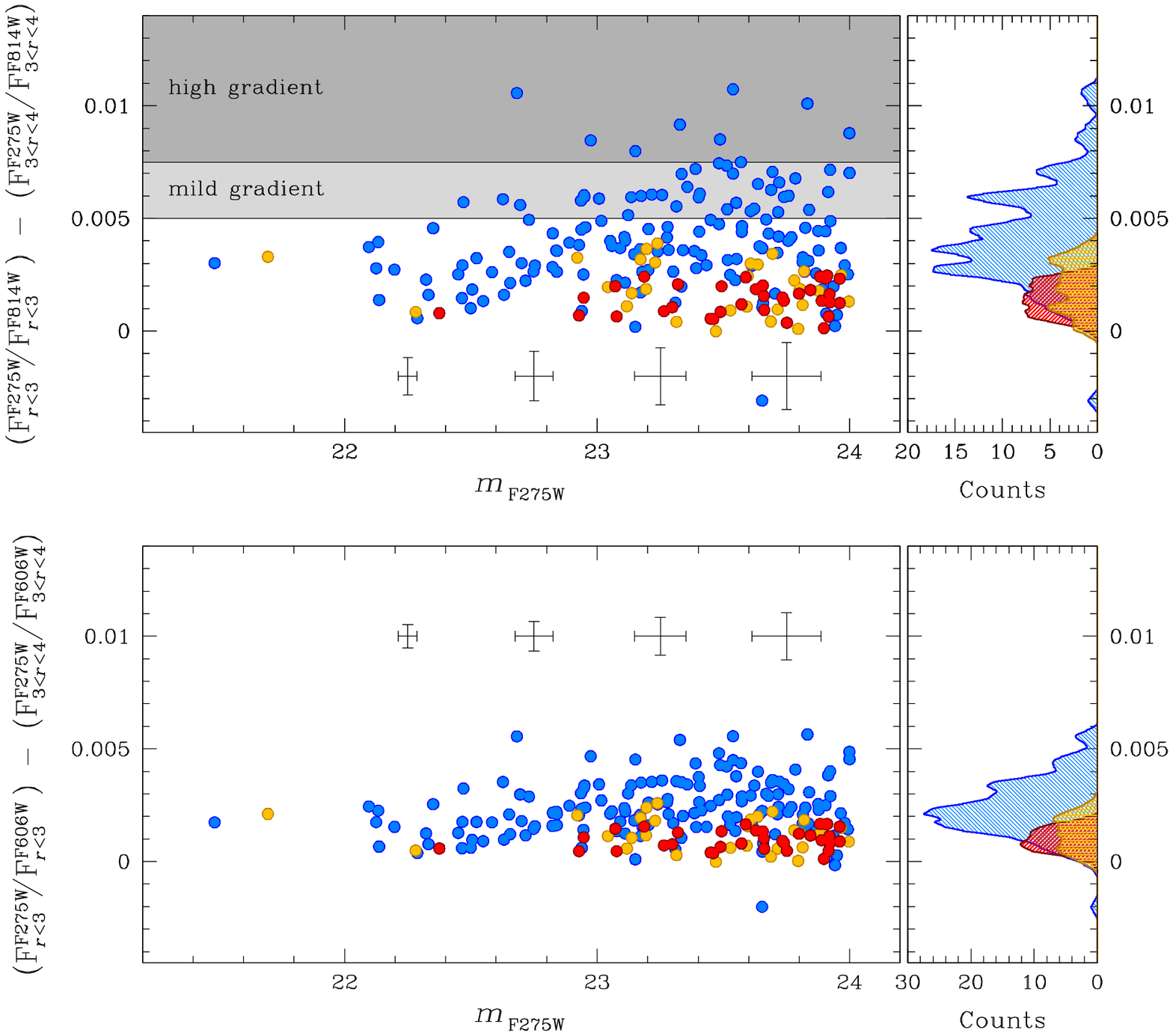}
\caption{The difference between the UV-to-optical flux ratios within
  the innermost 3 pixels and that between 3 an 4 pixels from the
  cluster centers, for the 214 brightest GCs in our M~87
  sample. Clusters are color-coded in blue, yellow and red if they
  belong to the bGC, iGC or rGC sequence, respectively. Error
    bars show the typical uncertainties for a given $m_{\rm F275W}$
    magnitude.  The corresponding color-gradient histograms are shown
  on the right-side of the figure.}
\label{f:color_gr}
\end{figure*}

\subsection{Mass estimates for the GCs in M87}
\label{ss:mass}
The right side of Figure~\ref{f:mw_ata} shows the total-mass scale for
the clusters. To estimate the GC masses we assumed the total mass of
$\omega$~Cen to be $4.05\times 10^{6}$ ${\rm M}_{\odot}$ (D'Souza \&
Rix 2013), and we linearly scaled this value with the F606W-band
luminosities, assuming the same mass-to-light ratio for all clusters.
This scale is just a rough estimate of the mass of MW and M~87
clusters, but it should represent a fair first-order mass ranking. For
example, the total mass of NGC~6397 is about $1.1\times 10^{5}$ ${\rm
  M}_{\odot}$ (Heyl et al. 2012), which is about what the scale
predicts over an order of magnitude away from the zero point. This
implies that at the faint limit (roughly at 10\% completeness level),
we can still detect in M~87 some GCs with masses as low as $\sim
3\times 10^{4}$ ${\rm M}_{\odot}$ like, e.g., NGC~6717 (Caloi \&
D'Antona 2011), while even less massive clusters are not included in
our sample.

Peng et al. (2009) showed that the brightest M~87 GCs have half-light
radii as large as 9 pc, corresponding to $\sim$5.8 pixels on our
stacks. A non-negligible fraction of the light of these clusters
necessarily falls outside our adopted 7-pixel-radius aperture
photometry. Using a 15-pixel-radius aperture photometry, the brightest
M~87 GCs are found to be $\sim 35$\% brighter than with the
7-pixel-based photometry, meaning that they might be up to $\sim 40\%$
more massive than what the mass scale in Figure~\ref{f:mw_ata} would
predict.

\begin{figure}[t!]
\centering
\includegraphics[width=\columnwidth]{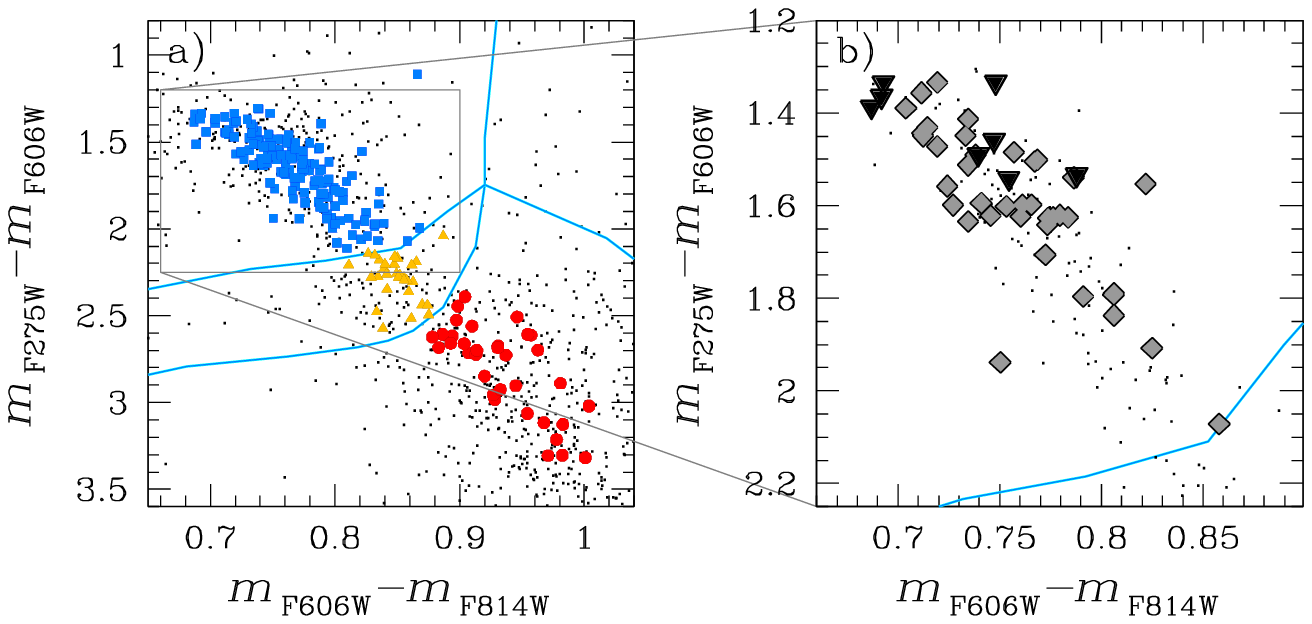}
\caption{Panel (a): two-color diagram of all M~87 GCs (in black), and
  of those used for the color-gradient analysis (in blue, yellow and
  red). A zoomed-in view is in panel (b), where only the analyzed GCs
  are shown (black dots). Clusters with a mild and a strong color
  gradient are shown as grey diamonds and black triangles,
  respectively.}
\label{f:color_gr1}
\end{figure}

\begin{figure*}[t!]
\centering
\includegraphics[width=\textwidth]{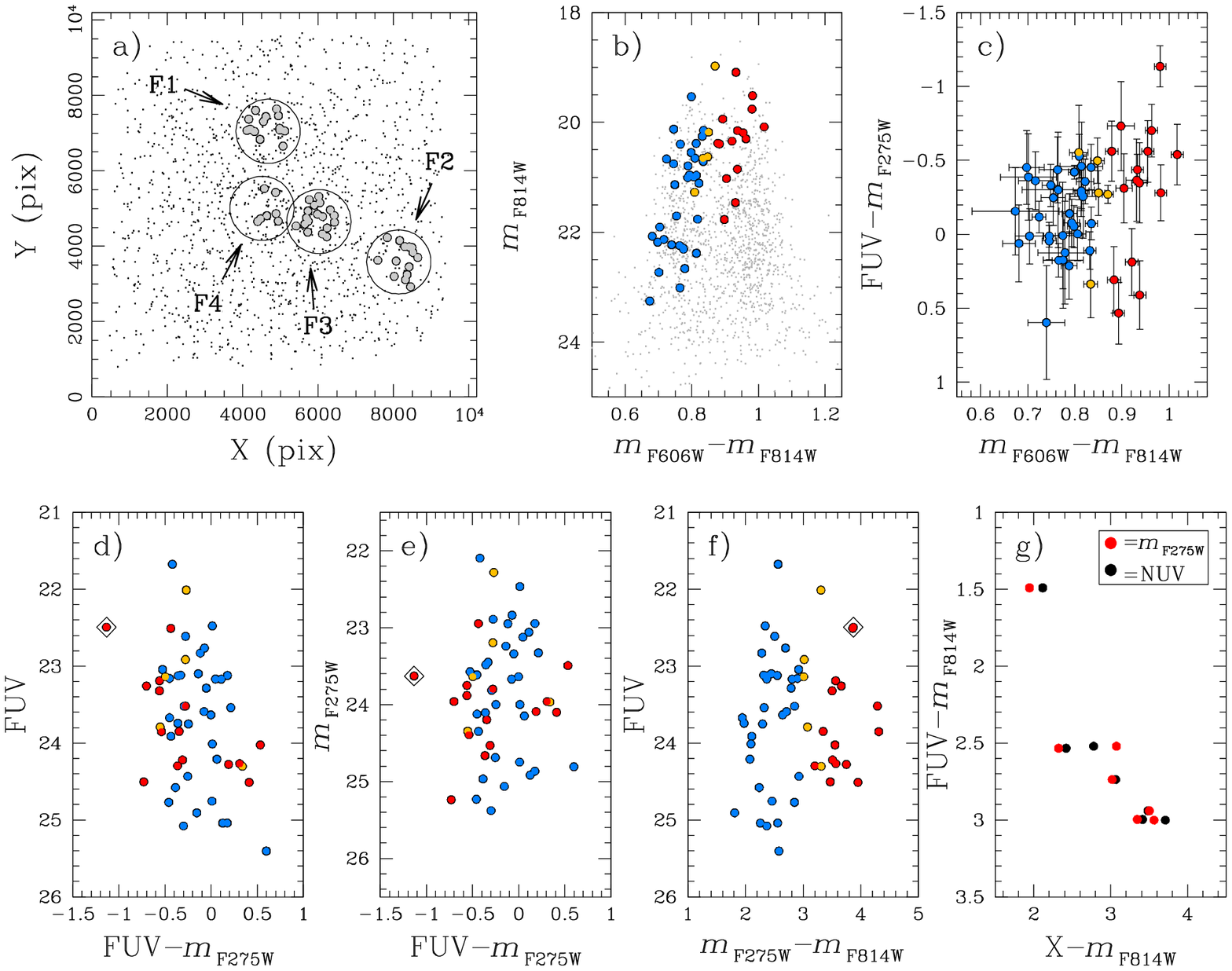}
\caption{Panel (a): Spatial distribution of S06 GCs cross-identified
  with our catalog (solid grey). The four fields imaged by S06 are
  highlighted.  (b) The UV-to-optical two-color diagram, with
  errorbars, for the common sources.  The bottom panels show three
  CMDs employing the FUV photometry. In panels (b) to (f) we
  color-coded each cross-identified GC according to its parent
  population. Panel (g) show the correlation between FUV and NUV
  photometry for common sources in the central field F4, as well as
  the good agreement between our F275W magnitudes and the NUV
  measurements of S06. The diamond seen in panels (d), (e) and (f)
  indicates the cluster with the bluest FUV-$m_{\rm F275W}$ color seen
  in panel (c).}
\label{f:tony}
\end{figure*}

\subsection{UV$-$optical colors and HB Morphology}
\label{sec:47}
The MW clusters for which we have extensive optical and UV photometry
allow us to undertake a simple experiment to quantify the sensitivity
of integrated UV$-$optical colors to the distribution of stars on the
HB. Figure~\ref{f:47tuc} shows the $m_{\rm F275W}$ vs.\ $m_{\rm F275W}
- m_{\rm F814W}$ CMD of $\omega$~Cen (NGC~5139, grey) and 47~Tuc
(NGC~104, black) with the HB stars of 47~Tuc being highlighted in
red. Various stellar groups and clumps are recognizable on the HB of
$\omega$~Cen, corresponding to the many sub-populations of this
cluster (see, e.g., Bellini et al.\ 2010), including the extreme blue
{\it hook} (see also Brown et al. 2010). Notice that the brightest
portion of the HB of $\omega$~Cen in the F275W band is not the blue
hook itself, but instead the next (redder) group of stars, with
$m_{\rm F275W} - m_{\rm F814W}\simeq 0$.  Thus, a maximum effect on
the integrated $m_{\rm F275W} - m_{\rm F814W}$ color would be produced
if all HB (red clump) stars of 47~Tuc were moved to this $m_{\rm
  F275W}$ brightest portion of the HB, as indicated by the red circle
in Figure~\ref{f:47tuc}.  By doing so, i.e., by moving all HB stars
from their actual location to the red circle, the synthetic colors of
47~Tuc become $\sim$ 0.01 mag bluer in $m_{\rm F606W} - m_{\rm F814W}$
and $\sim$ 0.98 mag bluer in $m_{\rm F275W} - m_{\rm F814W}$ with
respect to the untouched cluster. These new colors would place 47~Tuc
near the $m_{\rm F275W} - m_{\rm F814W}$ bluest rim of the M~87
distribution in the two-color plot of Figure~\ref{f:mw_piotto}.
The green arrow in Fig.~8d illustrates where 47~Tuc would be
  placed according to this shift. The new position on the two-color
  plot is very close to that of NGC~6717, a cluster with a blue HB and
  no red clump (see, e.g., Fig.~29 of Piotto et al.\ 2015). We
emphasize that these synthetic color shifts necessarily overestimate
the possible effect of secondary, helium-enriched populations, as
these may represent only a fraction, even if a major one, of the
entire stellar content of a GC.

Thus, a substantially smaller $m_{\rm F275W} - m_{\rm F814W}$ color
shift, of order of $\sim 0.5$ mag, is to be expected if --say-- only
$\sim$half of a cluster population were to populate the brightest
portion of the HB in the F275W passband. This is a shift comparable to
(or smaller than) the dispersion of the M~87 GC sequences, see panel
(c) of Figure~\ref{f:mw_piotto}.

A larger effect is to be expected when a bluer UV passband is
considered. We then repeated the same 47~Tuc experiment using the
F225W passband, with the result that the $m_{\rm F225W} - m_{\rm
  F814W}$ color shift is $\sim 1.86$ mag, hence this color is
substantially more sensitive to the presence of helium-enriched, blue
HB stars belonging to a cluster's second stellar generation.

\section{Color Gradients}
\label{sec:grad}

One important clue as to how the multiple populations may have
originated is given by the spatial distributions of the different
populations, which might retain information about their formation
sites. Among the massive MW GCs, the second generation is generally
more concentrated towards the center of the cluster compared to the
first generation (see, e.g., Bellini et al.\ 2009; Milone et
al.\ 2012a). As such, spatial gradients represent a fossil record of
the cluster's dynamical history. This finding has indeed promoted the
{\it cooling flow} model of D'Ercole et al. (2010, 2011). Over time,
the various stellar generations interact dynamically and may
homogenize their radial distribution in those clusters with short
relaxation time.

In the case of the Milky Way, only the most massive clusters (e.g.,
$\omega$~Cen, 47~Tuc, NGC~6441) have a half-mass relaxation time long
enough (greater than several Gyrs) to still retain some fossil record
of different formation sites of first- and second-generation stars.
Obviously, it is impossible to resolve M~87 GCs into stars and study
their spatial distribution star by star. However, helium-rich
second-population stars are bluer than He-normal stars. Thus, the
detection of a color gradient in the light profile of individual M~87
GCs would provide an indication for the presence in them of He-rich
sub-populations\footnote[9]{The color of HB stars is also ruled by a
  cluster's age, metallicity, mass, etc. The degeneracy with mass is
  broken by selecting the brightest M~87 clusters, while their
  metallicity can be inferred by the clusters' $V-I$ color. In this
  experiment we assume all M~87 clusters to have a similar age.}.

For this experiment we selected clusters brighter than $m_{\rm
  F606W}=22.7$ (corresponding to $\sim 5\times 10^{5}$ ${\rm
  M}_{\odot}$) and reasonably bright in UV ($m_{\rm F275W}<24.0$), so
to have a relatively long relaxation time and enough S/N.  A total of
214 M~87 GCs satisfied these selection criteria, specifically:\ 150
bGCs, 30 iGCs and 34 rGCs.  If we were to apply the same selection
criteria to MW GCs, we would have included just seven among the
Treasury GCs:\ NGC~2808, NGC~5024, $\omega$~Cen, NGC~5272, NGC~5286,
NGC~6715 and NGC~7089.

At the distance of M~87, 4.6 pc (the typical half-mass radius for the
selected GCs) corresponds to about 3 pixels on our stacks, and a
3-pixel-radius aperture photometry was measured for all the 214
selected sources. Typically, the flux of 29 pixels is summed within a
3-pixel radius (depending on the position of the source centroid
within the central pixel).  We then compared these
3-pixel-radius-based aperture fluxes with those measured in
the annulus between 3 and 4 pixels ($\sim 6.2$ pc) (enclosing 20
pixels), to keep similar Poisson statistics.

The flux ratios ${ {\rm F}^{\rm F275W}_{r<3}}/{{\rm F}^{\rm
    F814W}_{r<3} }$ within the 3-pixel radius were then computed and
compared with those between 3-pixel- and 4-pixel-radius apertures ${
  {\rm F}^{\rm F275W}_{3<r<4}}/{{\rm F}^{\rm F814W}_{3r<4} }$.  The
top-left panel of Figure~\ref{f:color_gr} shows these values as a
function of the $m_{\rm F275W}$ magnitude for bGCs, iGCs and rGCs (in
blue, yellow and red, respectively). iGCs and rGCs have very similar
colors within and outside the 3-pixel-radius apertures. bGCs, on the
contrary, have the same color-excess value just as lower limit, and
their color is in general appreciably bluer within 3 pixels than
outside, with some of them being up to 1\% bluer.  We highlighted 2
regions in the plot that define bGCs with a mild color gradient (38
GCs, in light grey) and with a higher color gradient (8 GCs, in dark
grey).  The histogram of the color-excess distribution for bGCs, iGCs
and rGCs is shown on the top-right panel (and color-coded
accordingly).

As a cross check, we repeated the same exercise using the F606W
passband instead of the F814W one. The results are shown in the lower
panels of Figure~\ref{f:color_gr}. The color differences are now
smaller than before, but the trend is still clear.  This UV-color
excess (or gradient) suggests that a good fraction of bGCs may indeed
be hosting He-rich second generation of stars that are still more
centrally concentrated than first-generation stars. Other UV-bright
sources, like blue-straggler stars, are not a valid explanation: while
on one hand they are expected to be more concentrated in the core of
GCs, nevertheless they are still 2--3 magnitudes fainter than blue HB
stars and substantially less numerous than HB stars among MW GCs
(e.g., Ferraro et al. 2003).

Finally, Figure~\ref{f:color_gr1} shows the location in the two-color
diagram of the clusters used for the color-gradient analysis. Panel
(a) shows all M~87 GCs with black points, whereas the 214 selected
clusters are color-coded according to their parent population (bGCs in
blue, iGCs in yellow, and rGCs in red). The lines used to divide the
clusters into subpopulations are also shown in light blue, for
clarity. A zoom-in view of the grey box in panel (a) is provided in
panel (b), where this time we are plotting only the selected clusters
(black dots). Clusters for which we found a mild color gradient are
highlighted with grey diamonds, while higher-color-gradient clusters
are marked by black triangles.  We also checked for the
  possibility of these color gradients being the result of passband
  differences in the shape of the PSFs. We found that PSF differences
  contribute less than $\sim 10\%$ to the ``high gradients'' seen in
  Figure~\ref{f:color_gr}.

We note that selected bGCs are spread all along the bGC sequence in
the two-color diagram of panel (a) of Figure~\ref{f:color_gr1}, but
those displaying a mild color gradient are preferentially found among
the bluer bGCs in both $m_{\rm F275W} - m_{\rm F606W}$ and $m_{\rm
  F606W} - m_{\rm F814W}$ colors. On the other hand, clusters with
high color gradients are only found among the bluest M~87 GCs. We
conclude that the GCs with both bluest $m_{\rm F275W} - m_{\rm F606W}$
color for their optical color and with a high color gradient represent
the best candidates in our sample for hosting helium-enriched
sub-populations producing blue HB stars.  Note that this is a
  preliminary exploration of color gradients in these M~87 GCs; a more
  quantitative analysis would require detailed profile fitting, which
  is beyond the purpose of the present paper.

\section{Comparison with previous FUV$-$NUV  studies}
\label{sec:tony}

As illustrated in Section~\ref{sec:47} and Figure~\ref{f:47tuc}, the
HB of helium-rich GC sub-populations can be very bright in the
ultraviolet, potentially making the clusters very blue in integrated
UV$-$optical colors. This is the case for the WFC3 UV filters we have
considered, and even more so for the STIS/MAMA FUV passband
(1150--1700 \AA), that samples substantially shorter wavelengths than
either the F225W and F275W passbands. FUV photometry should help us to
better characterize the possible presence of such extreme
helium-enriched sub-populations, as already been attempted by, e.g.,
S06 and K07 for M~87 GCs, and by Dalessandro et al. (2012) for MW
GCs. Here we want to take advantage of what we have learned so far
from our deep F275W, F606W and F814W integrated photometry and extend
our analysis to include the FUV. To this end, we cross-identified
sources in common between our catalogs and those published in S06 and
Dalessandro et al. (2012).

Among the 66 M~87 sources with reliable FUV photometry in S06, only 53
are in common with our catalog. A visual inspection of the missing
sources in our stacks revealed that they are either background
galaxies (having a distorted shape in at least one of our stacks), are
not found in F814W, or have non-positive flux in
F275W. \footnote[10]{We did not convert the FUV STMAG values to the
  VEGAMAG system because of the lack of a proper photometric
  conversion between the two photometric systems. This introduces a
  zero-point offset that is the same for all GCs.  Since we are
  interested here in finding objects that have an anomalous behavior
  with respect to the average trend, this zero-point offset is of no
  consequence for our analysis.}

\begin{figure}[t!]
\centering
\includegraphics[width=\columnwidth]{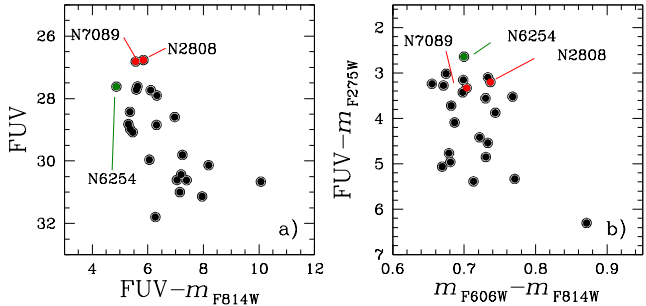}
\caption{FUV vs. FUV$-m_{\rm F814W}$ CMD (a) and two-color
  diagram (b) of MW GCs obtained by cross-identifying our MW
  catalog with the GALEX FUV photometry of Dalessandro et
  al. (2012). The latter compares to panel (c) of Figure~\ref{f:tony},
  for M~87 GCs.}
\label{f:dale}
\end{figure}

Panel (a) of Figure~\ref{f:tony} shows the FoV of our M~87 field, with
all the reliable sources in our catalog as black dots. The four fields
imaged by STIS/MAMA are tagged and highlighted with circles. The 53
sources in common with S06 are shown as filled grey circles. Panel (b)
shows the optical CMD of all our M~87 GCs (grey dots), and of those
cross-identified with S06, color-coded according to their parent
population (bGC, iGC or rGC). The two-color diagram employing FUV
photometry is presented in panel (c), with photometric error bars.
Panels (d), (e) and (f) show three CMDs based on combinations of FUV,
WFC3/UVIS and ACS photometry.  There is marginal evidence that among
these 53 common objects the brightest rGCs are also the bluest in the
FUV$-m_{\rm F275W}$ color, with one cluster (highlighted with an open
diamond in panels (d), (e) and (f) standing out as the bluest of all.
The fact that this cluster is simultaneously red in the optical color
and blue in the UV color suggests that it might host two
populations:\ a metal-rich first generation and a He-rich second
generation with a very hot HB. As such, it could represent an extreme
analog of NGC~6388 and NGC~6441 in the MW.  Note that this object is
still red when the UV$-$optical color is based on the F275W filter
(panel f), indicating that its HB may be mostly occupied by
extremely-hot stars (blue-hook stars in Figure 10).  In any case,
objects like this one seem to be the exception rather than the rule (1
out of 53). Perhaps more interesting is the fact that most rGCs
  (and most iGCs as well) are bluer than bGCs in the FUV$-m_{\rm
    F275W}$ color, whereas they are redder in all other colors. This
  may signal that indeed these clusters harbor many blue-hook stars.

The central field of S06 (F4 in panel a) also contains photometry in
the NUV passband.  We have 7 GCs in common between our catalog and
that of S06 in this field. Panel (g) of Figure~\ref{f:tony} shows the
correlation between the FUV$-m_{\rm F814W}$ and X$-m_{\rm F814W}$
colors, where X is either F275W (in red) or NUV (in black). It is
worth mentioning the good agreement between our F275W photometry and
S06 NUV measurements, despite the different adopted photometric
system.

What about our Galaxy? Does the MW have similar GCs compared to the
anomalous ones we found in M~87? To find this out, we cross-identified
our MW catalog with the integrated GALEX FUV photometry published in
Dalessandro et al. (2012).  We did not convert their ABMAG values into
VEGAMAG, for the same reasons stated above in the case of S06 STMAG.

There are 23 clusters in common between Dalessandro et al. (2012) and
our 45 MW GCs discussed in Section~\ref{sec:mw}.
In Figure~\ref{f:dale} we show the FUV vs. FUV$-m_{F814W}$ CMD on the
left (panel a), and the UV-to-optical two-color diagram on the right
(panel b). All magnitudes are on the same reference system of M~87 GCs
(i.e., the same distance and reddening of M~87, but \textit{not} the
same photometric system). The latter diagram should be compared to
panel (c) of Figure~\ref{f:tony} for M~87 GCs.

The two brightest objects in FUV are, not surprisingly, NGC~2808 and
NGC~7089 (M2) in red in both panels\footnote[11]{The Dalessandro et
  al. catalog does not provide the FUV photometry for $\omega$~Cen.}.
The former cluster has at least three sub-populations with He
abundances up to $Y\simeq 0.4$ (Piotto et al. 2007). The latter hosts
seven subpopulations with He content up to 0.33, with two
sub-populations making $\sim 4\%$ of the cluster being substantially
enriched also in iron as well (Milone et al. 2015). Both clusters have
a well populated and extended blue HB, with multiple clumps. Yet, they
do not stand out for having particularly blue FUV$-m_{F275W}$ or
$m_{F275W}-m_{F814W}$ colors.  The bluest object in panel (a) of
Figure~\ref{f:dale} is NGC~6254 (M~10), (marked in green in both
panels). This object is also the bluest in FUV$-m_{\rm F275W}$
color. NGC~6254 has a well-populated BHB (see, e.g., Figure~5 of
Dalessandro et al. 2013), and the presence of distinct multiple
populations within this clusters has just been reported (Piotto et
al. 2015).

At odds with M~87 GCs, we find no objects simultaneously red in the
optical color and blue in the FUV-based color among these 23
clusters. Moreover, clusters that are red in the optical have about
the same UV color (or are even redder) in FUV$-m_{\rm F275W}$ (panel
b).

\section{Discussion}
\label{sec:discu}

We have imaged through the UV F275W passband the central
$2\farcm7\times 2\farcm7$ field of the giant elliptical galaxy M~87,
using the WFC3/UVIS camera onboard the \textit{HST}. In combination
with archival ACS/WFC F606W and F814W data covering the same field, we
have constructed various optical and UV$-$optical color-magnitude and
two-color plots in an attempt to identify candidate globular clusters
hosting a significant helium-enriched sub-population, analogous to
those present in several GCs of our own Galaxy, that produce very-blue
HB stars. This experiment has attained only partial success. Out of
the 1460 candidate GCs identified in all three bands, the best
candidates for hosting helium-enriched sub-populations consist of a
small group of clusters that are both slightly bluer in $m_{\rm F275W}
- m_{\rm F606W}$ for their optical color than MW GCs, and exhibit a UV
color gradient.

Using real stellar data for some MW globulars, we have empirically
estimated the maximum possible effect on UV$-$optical colors of a
secondary, helium-enriched population. Assuming that such a population
makes 50\% of the cluster mass, the maximum blueing of the cluster
integrated color is $\sim 0.5$ mag in $m_{\rm F275W} - m_{\rm F814W}$
and $\sim 1$ mag in $m_{\rm F225W} - m_{\rm F814W}$. This $m_{\rm
  F275W} - m_{\rm F814W}$ shift is similar to or smaller than the
color width of either the blue or the red GC sequence in M~87, hence
it does not allow us to unambiguously detect clusters with an
exceptionally blue HB. Clearly, the $m_{\rm F225W} - m_{\rm F814W}$
color would offer a better chance to do this, but WFC3/UVIS exposures
with the bluer F225W filter would have required considerably longer
integration times.

For a set of 45 GCs in the MW we have constructed integrated
$m_{F275W}-m_{F814W}$, $m_{F275W}-m_{F606W}$ and $m_{F606W}-m_{F814W}$
colors, directly from deep photometry of individual stars that are
members of the clusters. After reddening corrections, the UV$-$optical
colors of these MW clusters are compared to those of M~87.  We
extended the comparison to 53 Milky-Way GCs (but only to F606W and
F814W magnitudes), and used them to derive a first-guess estimate of
the mass of M~87 GCs.

We included FUV information for 53 M~87 GCs from S06, and found that
the brightest rGCs are also as blue as --or even bluer-- than the
brightest bGCs in FUV$-m_{\rm F275W}$, with one object being
significantly bluer. A similar comparison is made also for 23 MW GCs
using Dalessandro et al. (2012) FUV integrated magnitudes. In this
case, the bright MW red clusters in optical are not as blue as the
bright blue cluster in neither of the colors employing the FUV, at odd
with M~87 clusters. Unfortunately, the lack of a
statistically-significant sample of metal-rich MW clusters with FUV
photometry does not allow us to make a full comparison with the M~87
GC system.

Other results of this investigation include:
\begin{itemize}
\item{Both red and blue GCs appear to have a slightly flattened
    distribution, nearly perpendicular to the famous jet, with this
    flattening being more pronounced for the blue GCs.}
\item{We find that M~87 blue GCs span the same UV$-$optical color
    range as blue GCs in the MW. However, they appear to be redder by
    $\sim 0.06$ mag in the optical color $m_{\rm F606W} - m_{\rm
      F814W}$, compared to MW GCs, producing a systematic relative
    shift in the two-color diagram in qualitative agreement with
    earlier findings by Sohn et al. (2006) and Kaviraj et
    al. (2007). However, we cannot exclude that this color shift may
    result from differences in how colors are measured in the two
    cluster families.}
\item{Only a handful of M~87 GCs in this survey field appear to be
  more massive than $\omega$~Cen, the most massive cluster in the MW,
  with some as massive as a few $10^7$ M$_\odot$.}
\item{We identify a small sample of M~87 red GCs which are extremely
  blue in FUV$-m_{\rm F275W}$ (Fig.~\ref{f:tony}c), which have no
  known counterparts among the MW GCs (Fig.~\ref{f:dale}).  Such blue
  colors suggests they may contain a unusual number of extremely blue,
  helium rich HB stars.}
\item{As a side product, in an Appendix we  briefly
  discuss the images of the jet, counterjet and dust lanes near the
  center of M~87, which become especially evident in the ultraviolet
  F275W stack.}
\end{itemize}

We release the photometric catalog of M~87 GCs in the F275W, F606W and
F814W bands to the astronomical community, as well as the three
high-quality image stacks in the same filters.

\acknowledgments \noindent \textbf{Acknowledgments.} The authors
gratefully thank the anonymous referee for his/her useful suggestions
that helped improving the manuscript, and Thomas Puzia for a critical
reading of the manuscript. AB and JA acknowledge support from STScI
grant GO-12989. APM acknowledges the financial support from the
Australian Research Council through Discovery Project grant
DP120100475. MS acknowledges support from STScI grant GO-13297 and
Becas Chile de Postdoctorado en el Extranjero project 74150088.

\appendix

\section{Jet, Counterjet and Dark Lanes}

\begin{figure*}[b!]
\centering
\includegraphics[width=17cm,angle=-90]{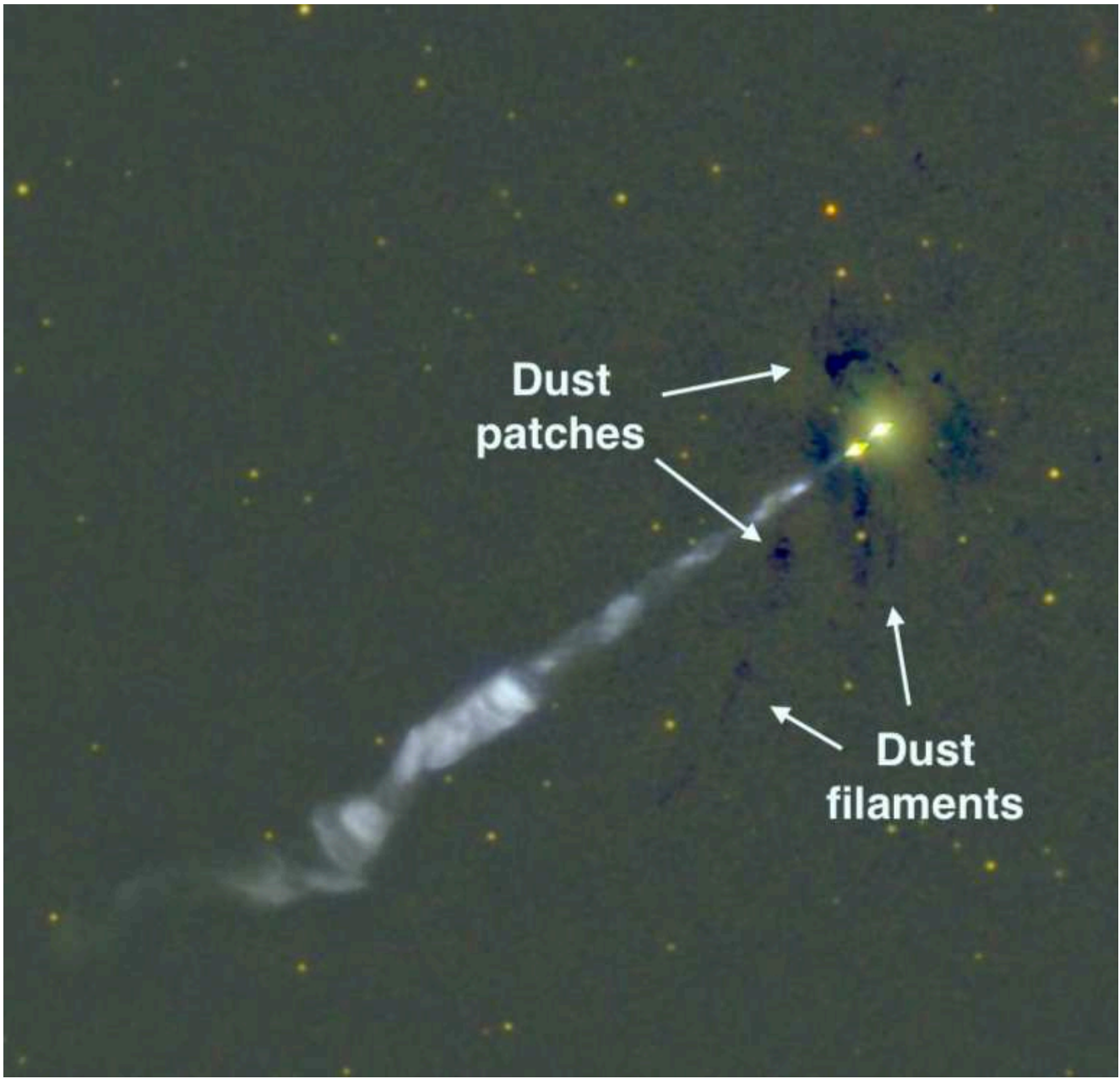}
\caption{A zoomed-in view of the region around the center of the
  galaxy and the jet. The color scale has been stretched to enhance
  filaments and patches of dust.}
\label{f:app}
\end{figure*}

We briefly comment here on some of the features related to the
prominent jet , so evident in the stack images.  We may return to some
of these issues in the future, but for this paper we are primarily
focused on the GC population.

Besides the famous jet of M~87, one obvious counterjet feature is
clearly visible in Figure~1, with three other faint filamentary
structures being possibly related to the counterjet (such as
counterjet-generated shock fronts).  The red-green-blue color-scale of
this panel is also enhancing the visibility of these dim
features. Some of them are bright in F606W (marked as ``Green
  features'' in the figure), while others are almost invisible in
this band, yet are still bright in the other two bands
(``Purple features''). These color differences may be due to
different emission-line ratios resulting from different excitations
among the features.

The dark areas near the center of M~87 visible in Figure~1 are not an
artifact of the galaxy-subtraction routines but are real features in
the scene. Figure~\ref{f:app} shows these features in more detail,
having been optimized for this purpose.  Various dust filaments around
the galactic center and the jet are now clearly visible: they are also
present in the non sky-subtracted stacks, although they stand out less
clearly, due to the scattered light of the galactic core. They are
visible in all three bands, but are definitely more prominent in the
F275W image, as expected from obscuring dust clouds. Notice also the
sharp edges of some of them, with one being clearly in front of the
jet and attenuating the light of its obscured section, suggesting they
could be the swan song of recent tidally-disrupted dwarf galaxies.

The phenomenology of the M~87 core is typical of what one expects to
happen in a very massive elliptical galaxy. Supernova heating at the
known rate of Type Ia supernov\ae\ is indeed insufficient to prevent
gas outflows to eventually turn to inflows following a central-cooling
catastrophe:\ as stellar mass loss accumulates, the hot gas density
increases until cooling exceeds the supernova heating (Ciotti et
al.\ 1991). Under such circumstances, the inflowing material can feed
the central supermassive black hole, starting a series of intermittent
nuclear activity with duty cycles of several $10^8$ yrs (Ciotti \&
Ostriker 2001; Novak, Ostriker \& Ciotti 2012). This kind of recurrent
activity is indeed instrumental in keeping the galaxy quenched, once
star formation has been quenched for the first time (by whatever
mechanism).

Coming back to the dark spots, one interesting question concerns the
origin of the obscuring dust. Forming and growing dust in the
interstellar medium itself, at the time of the cooling catastrophe
near the center, is one option. However, the density may not be high
enough for rapid nucleation except perhaps near the black hole
itself. Alternatively, dust may form in the wind of mass-losing red
giants and AGB stars and survive sputtering just in the cooler gas
resulting from the cooling catastrophe. We have no decisive argument
either way, other than that these dust features are most probably
vestiges of the events that lead to the re-fueling of the central
black hole.

One can also notice that bright regions appear in between dark
patches.  Their intrinsic nature may actually be similar to the dark
patches, possibly being dust-rich clouds that shine of dust-scattered
light from the AGN itself.

\end{document}